\documentclass{aa}

\usepackage{graphics}
\newcommand\aaps{A\&AS}
\newcommand\apj{ApJ}
\newcommand\apjl{ApJL}
\begin{document}

%
%#########################################
%                title
%#########################################
%
\title {
Helium-like triplet density diagnostics:
}
\subtitle{Applications to CHANDRA--LETGS X-ray 
observations of Capella and Procyon
}
\author{
        J.-U. Ness\inst{1}
        \and
        R. Mewe\inst{2}
        \and
        J.H.M.M. Schmitt\inst{1}
        \and
        A.J.J. Raassen\inst{2,3}
        \and
        D. Porquet\inst{4}
        \and
        J.S. Kaastra\inst{2}
        \and
        R.L.J. van der Meer\inst{2}
        \and
        V. Burwitz\inst{5}
        \and
        P. Predehl\inst{5}
        }

\institute{
 Universit\"at Hamburg, Gojenbergsweg 112, D-21029 Hamburg, Germany
 \and
 Space Research Organization Netherlands (SRON),
 Sorbonnelaan 2, 3584 CA Utrecht, The Netherlands
  \and
 Astronomical Institute "Anton Pannekoek", Kruislaan 403,
 1098 SJ Amsterdam, The Netherlands
 \and
 CEA/DSM/DAPNIA, Service d'Astrophysique, CEA Saclay, F-91191 Gif sur Yvette Cedex, France
  \and
  Max-Planck-Institut f\"ur Extraterrestrische Physik (MPE), Postfach 1603,
  D-85740 Garching, Germany
}

\authorrunning{J.U. Ness, R. Mewe, J.H.M.M. Schmitt et al.}

\titlerunning{Coronal densities for Capella and Procyon}

\offprints{J.-U.\ Ness}
\mail{jness@hs.uni-hamburg.de}
\date{received \today; accepted December 01, 2000}
\thesaurus {02(02.01.3), 03(03.20.8), 08(08.09.2 Capella,08.09.2 Procyon, 08.03.5, 08.12.1, 08.01.2), 13(13.25.5)} 
\maketitle

%#########################################
%                abstract
%#########################################
%
\begin{abstract}
Electron density diagnostics based on the triplets of Helium-like C\,{\sc v}, N\,{\sc vi},
and O\,{\sc vii} are applied to the X-ray spectra of Capella and Procyon measured with
the Low Energy Transmission Grating Spectrometer (LETGS) on board 
the Chandra X-ray Observatory.
New theoretical models for the calculation of the line ratios between the
forbidden (f), intercombination (i), and the resonance (r) lines of the
helium-like triplets are used. The (logarithmic) electron densities
(in cgs units) derived from the f/i ratios for Capella are
$<~9.38\,{\rm cm^{-3}}$ for O\,{\sc vii} (2$\sigma$ upper limit)
(f/i=4.0$\pm 0.25$),
$9.86\pm0.12~{\rm cm^{-3}}$ for N\,{\sc vi} (f/i=1.78$\pm 0.25$),
and $9.42\pm0.21~{\rm cm^{-3}}$
for C\,{\sc v} (f/i=1.48$\pm 0.34$), while for Procyon we obtain
$9.28^{+0.4}_{-9.28}~{\rm cm^{-3}}$ for O\,{\sc vii} (f/i=3.28$\pm 0.3$),
$9.96\pm0.23~{\rm cm^{-3}}$ for N\,{\sc vi} (f/i=1.33$\pm 0.28$),
and $<8.92$~${\rm cm^{-3}}$ for C\,{\sc v} (f/i=0.48$\pm 0.12$).
These densities are quite typical of densities found
in the solar active regions, and also pressures and temperatures
in Procyon's and Capella's corona at a level of $T\sim10^6$ K are quite 
similar. We find no evidence for densities as high
as measured in solar flares. Comparison of our Capella and Procyon 
measurements with the Sun shows little difference in the physical properties
of the layers producing the C\,{\sc v}, N\,{\sc vi}, and O\,{\sc vii} emission.
Assuming the X-ray emitting plasma to be confined in magnetic loops, we
obtain typical loop length scales of ${L_{Capella}} \ge 8~{L_{Procyon}}$
from the loop scaling laws, implying that the magnetic structures
in Procyon and Capella are quite different.
The total mean surface fluxes emitted in the helium- and
hydrogen-like ions are quite similar for Capella and Procyon, but
exceed typical solar values by one order of magnitude.
We thus conclude that Procyon's and Capella's coronal filling factors
are larger than corresponding solar values.

\end{abstract}
%
%#######################################################
%                  Introduction
%#######################################################

\section{Introduction}

The hot plasma in the corona of the Sun and of other stars is thought to be
in ``coronal equilibrium''. Atomic excitations occur through collisions
with electrons. The excited atoms decay radiatively and the emitted
radiation escapes without any further interaction with the emitting plasma.
As a consequence, the emission is optically thin, and the total flux
emitted in some spectral band or in a given emission line is proportional
to the emission measure $EM$, defined as the integral
of the square of the plasma density $n$ over the emitting volume elements
d$V$ through $ EM = \int n^2 {\rm d}V $.
Thus, observationally, the contributions of
density and volume to a given observed value of $EM$ cannot be disentangled.

Stellar X-ray surveys carried out with the {\it Einstein} and ROSAT satellites
have shown an enormous range of X-ray luminosity ($L_{\rm X}$)
for stars of given spectral type (cf., \cite{vaiana81}, \cite{schmitt97}).
Typically, one observes star to star variations in $L_{\rm X}$
of up to four orders
of magnitude, with the largest X-ray luminosities found among the stars
with the largest rotation rates. While one definitely finds a correlation
between mean coronal temperature and X-ray luminosity
(\cite{schmitt85}; \cite{schmitt97}), it is also clear that the single-most
important factor contributing to the large variations in $L_{\rm X}$ is the
variation in emission measure. The conclusion therefore is that active stars
(can) have a couple of orders of magnitude higher coronal emission measure,
while maintaining the same optical output as low-activity stars like our Sun.

The emission measure is directly linked to the structure of stellar coronae,
if we assume, going along with the solar analogy, that the X-ray emitting
plasma of a stellar corona is confined in magnetic loops.
The observed values of $EM$ and $L_{\rm X}$ for a given star could
be accounted for either by the existence of
more loops than typically visible on the solar surface, by higher density
loops or by longer, more voluminous loops. Thus the question is reduced to the
following: if $EM_{\star} \gg EM_{\odot}$ for an active 
star, one wants to know whether this is due to $n_{\star} > n_{\odot}$ or
$V_{\star} > V_{\odot}$ or both.

Spatially resolved solar observations allow to disentangle density and volume
contributions to the overall emission measure.
One finds the total X-ray output of the Sun dominated -- at least under maximum
conditions -- by the emission from rather small, dense loops.
Stellar coronae always appear as point sources. The only way to infer
structural information in these unresolved point sources has been via
eclipse studies in suitably chosen systems where one tries
to constrain the emitting plasma volume from the observed light curve. 
Another method to infer structure in spatially unresolved data 
are spectroscopic measurements of density. 
The emissivity
of plasma in coronal equilibrium in carefully selected lines does depend
on density.
 Some lines may be present in low-density plasmas and
disappear in high-density plasmas such as the forbidden lines
in He-like triplets, while other lines may appear in high-density
plasmas and be absent in low-density plasmas (such as lines formed following
excitations from excited levels).
With the high-resolution spectroscopic facilities onboard {\it Chandra} it is
possible to carry out such studies for a wide range of X-ray sources.
The purpose of this paper is to present and discuss some key density
diagnostics
available in the high-resolution grating spectra obtained with
{\it Chandra}. We will specifically discuss the spectra obtained
with the Low Energy Transmission Grating Spectrometer (LETGS)
for the stars Capella and Procyon.\\

\begin{table}
\caption[ ]{\label{star_prop}Properties of Procyon and Capella:
mass $M$, radius $R$, effective temperature $T_{\rm eff}$, log~g and
the limb darkening coefficient $\epsilon$}
\renewcommand{\arraystretch}{1.2}
\begin{tabular}{r r r}
\hline
&Procyon&Capella\\
\hline
d/pc&3.5&13\\
$M/M_\odot$&$1.7\pm0.1^4$&$2.56\pm0.04^3$\\
$R/R_\odot$&$2.06\pm0.03^4$&$9.2\pm0.4^3$\\
$T_{\rm eff}$/K&$6530\pm 90^2$&$5700\pm 100^3$\\
log g&$4.05\pm0.04^2$&$2.6\pm0.2^5$\\
Spectr. type & F5~IV-V & Ab: G1~III\\
&& (Aa: G8/K0~III)\\
$\epsilon$&0.724$^1$&0.83$^1$\\
\hline
\end{tabular}\\
{References:\\
$^1$\cite{diaz95},\\
$^2$\cite{fuhrm97},\\
$^3$\cite{hummel94},\\
$^4$\cite{irwin92},\\
$^5$\cite{kelch78}}
\begin{flushleft}
\renewcommand{\arraystretch}{1}
\end{flushleft}
\end{table}

Both Capella and Procyon are known to be relative steady and strong X-ray
sources; no signatures of flares from these stars
have ever been reported in the literature.
Both Capella and Procyon are rather close to the Sun at distances
of 13~pc and 3.5~pc (Tab.~\ref{star_prop}),
so that effects of interstellar absorption are very
small. Both of them have been observed with virtually all
X-ray satellites flown so far. Capella was first detected
as an X-ray source by \cite{catura75}, and confirmed by \cite{mewe75}, Procyon by \cite{schmitt85}.
The best coronal spectra of Capella were obtained with the
{\it Einstein} Observatory FPCS and OGS (\cite{vedder83}, \cite{mewe82}), the
EXOSAT transmission grating (\cite{mewe86}, \cite{lemen89}) and EUVE
(\cite{dupree93}, \cite{schrijver95}),
while high-spectral resolution spectral 
data for Procyon have been presented by \cite{mewe86} and \cite{lemen89} using EXOSAT
transmission grating data and \cite{drake95} and \cite{schrijver95} using EUVE data. Note that
\cite{schmitt96b} and \cite{schrijver95} investigated the coronal density of Procyon using
a variety of density sensitive lines from Fe\,{\sc x} to Fe\,{\sc xiv} in the EUV range
and found Procyon's coronal density consistent with that of solar active
region densities.\\
The plan of our paper is as follows: We first briefly review the atomic
physics of He-like ions as applicable to solar (and our stellar)
X-ray spectra. We briefly describe the spectrometer used to obtain
our data, and discuss in quite some detail the specific procedures
used in the data analysis, since we plan to use these methods in all
our subsequent work on {\it Chandra} and XMM-Newton spectra. We then proceed
to analyze the extracted spectra and describe in detail how we dealt 
with the special problem of line blending with higher dispersion orders. Before
presenting our results we estimate the formation temperatures of the lines, the
influence of the stellar radiation field and the influence of optical depth
effects followed by detailed interpretation. The results
will then be compared to measurements of the Sun and we 
close with our conclusions.

\section{Atomic physics of He-like ions}

The theory of the atomic physics of helium-like triplets has been
extensively described in the literature
(\cite{gabriel69}, \cite{blume72}, \cite{mewe78}, \cite{pradhan81a},
\cite{pradhan81b}, \cite{pradhan82}, \cite{pradhan85}, and recently
\cite{porquet00} and \cite{mewe00}).
Basically, the excited states (1s2{\it l}) split up into the terms $2 ^1P$,
$2 ^3P,2 ^1S$, and $2 ^3S$, out of which the levels with J $\ne$ 0 decay to the
ground state $1 ^1S$ through the 
resonance line (abbreviated by r), the intercombination line
(i) and the forbidden line (f), respectively; the latter two lines
involve spin changes and therefore violate the selection
rules for electric dipole radiation.
Although the radiative transition rate for the forbidden line
is quite small, in a low-density plasma collisional
depopulation processes are so rare, that
the excited $2 ^3 S$ state does
decay radiatively. In a high-density plasma collisional deexcitations
dominate and hence the forbidden line disappears.
Complications arise from other competing processes populating and depopulating
the $^3P$ and $^3S$ levels. These are in particular radiative transitions
induced by the underlying photospheric stellar radiation field (discussed in
Sect.~\ref{rad_field}) as well as ionization and recombination processes from
the Li-like and H-like ions.

It is customary to describe the measured line intensities r, i and f in 
terms of the ratios
\begin{equation}
\label{ratios}
R_{\rm obs} = \frac {\rm f} {\rm i}\\
{\rm and}\\
G_{\rm obs} = \frac {\rm i + f} {\rm r}\,.
\end{equation}
In this paper we use the notation used in e.g. \cite{pradhan81b}. In other
contexts (e.g. calculating ions of higher ionization stages) another notation
labeling r as {\bf w}, f as {\bf z} and i as {\bf x+y}
is often used pointing out that the intercombination
line separates via M2 and E1 transitions into two components.
Theory describes the density sensitivity of the R ratio by the functional 
dependence
\begin{equation}
\label{R_Ne}
R(N_{\rm e}) \ = \ R_{\rm 0} \frac {1} {1 + \phi/\phi_{\rm c} + N_{\rm e}/N_{\rm c}}\,.
\end{equation}
Densities are inferred from equating $R(N_{\rm e})$ with $R_{\rm obs}$.
For convenience, we follow here the expressions as given by \cite{blume72}.
The low-density limit $R_{\rm 0}$, which applies for $N_{\rm e} = 0$ and
$\phi = 0$, is given by
\begin{equation}
R_{\rm 0}=\frac{1+F}{B}-1,
\end{equation}
with the radiative branching parameter $B$ averaged over the two
components of the intercombination line.
The parameter $N_{\rm c}$, the so-called critical density (cf.,
Tab.~\ref{ionvals}), above which the observed line ratio is density-sensitive,
is given by
\begin{equation}
N_{\rm c} = \frac{A(2 ^3S_1 \rightarrow\ 1 ^1S_0)}{(1+F)C(2 ^3S_1 \rightarrow\ 2 ^3P)}\,, 
\end{equation}
where ${A(2 ^3S_1 \rightarrow\ 1 ^1S_0)}$ is the radiative transition probability of the forbidden line,
$C(2 ^3S_1 \rightarrow\ 2 ^3P)$ is the electron collisional rate coefficient
 for the transition from $2 ^3S$ to $2 ^3P$,
and $F$ is the ratio of the collision rates from the
 ground to the levels $2 ^3S$ and $2 ^3P$, respectively
(including cascade effects from upper levels through radiative transitions 
after excitation or recombination).
Finally, the parameters $\phi$ (the radiative absorption rate from $2 ^3S$ to
$2 ^3P$ induced by an external 
radiation field) and $\phi_{\rm c}$ describe the additional
possible influence of the stellar radiation field on the depopulation of the
$^3 S$ state (cf., Sect.~\ref{rad_field}). The latter is insignificant
especially for the He-like triplets of higher ionization stages than C\,{\sc v}.
Values for $R_{\rm 0}$ and $N_{\rm c}$ used in this paper are listed in
Tab.~\ref{ionvals}.
Obviously, for very large densities or large radiation fields the forbidden
line and hence $R$ disappears. The ratio
\begin{equation}
G=(i+f)/r,
\end{equation}
a measure of the relative strength
of the resonance line, depends on the electron temperature, $T_{\rm e}$
(e.g., \cite{mewe00}).
$G$ can be used to derive temperatures for the lines used for
density diagnostics that can be compared to the maximum formation temperature
$T_{\rm m}$.

\begin{table}
\caption[ ]{\label{ionvals}Atomic Parameters for He-like triplets.
$T_{\rm m}$ is the peak line formation temperature (MEKAL), 
$R_{\rm 0}$ is the low-density limit and $N_{\rm c}$ the density
were $R$ falls to half its low-density value.}
\begin{flushleft}
\renewcommand{\arraystretch}{1.2}
\begin{tabular}{r r r r}
\hline
ion & $T_{\rm m}$/MK & $R_{\rm 0}$ & $N_{\rm c}/(10^{10}cm^{-3})$\\
\hline
C\,{\sc v}   &   1          & 10.6        &   0.051\\
\hline
N\,{\sc vi}  &   1.4        & 4.9         &   0.45\\
\hline
O\,{\sc vii} &   2.0        & 3.495       &   3.00\\
\hline
\end{tabular}
\renewcommand{\arraystretch}{1}
\end{flushleft}
\end{table}

%#######################################################
%                 Technical Details 
%#######################################################
\section{Instrument description}

The Low Energy Transmission Grating Spectrometer (LETGS) on board the Chandra Observatory
is a diffraction grating spectrometer covering the wavelength range between
2\,--\,175~\AA\ (0.07\,--\,6~keV). 540 individual grating elements are mounted
onto a toroidal ring structure. Each of the elements consists of a 
freestanding 
gold grating with 1$\mu$m grating period. The fine gold wires are held by two
different support structures, a linear grid with 25.4~$\mu$m and a coarse
triangular mesh
with 2~mm spacing. The whole grating ring can be inserted into the convergent
beam just behind the High Resolution Mirror Assembly (HRMA) thereby dispersing
the light of any X-ray source in the field of view into its spectrum.
The efficiency of the grating
spectrometer is of the order of 10\% on average but is enhanced by a
factor of two around 2~keV due to partial transparency effects;
a more detailed description of the instrument is presented by \cite{predehl97}.
Both sides of the spectrum are recorded with a microchannel plate detector
(HRCS), placed behind the transmission grating. In contrast to CCD based
detectors, the HRCS detector provides essentially no intrinsic energy
resolution, the energy information for individually recorded events is
solely contained in the events' spatial location.

\begin{figure}
 \resizebox{\hsize}{!}{\includegraphics{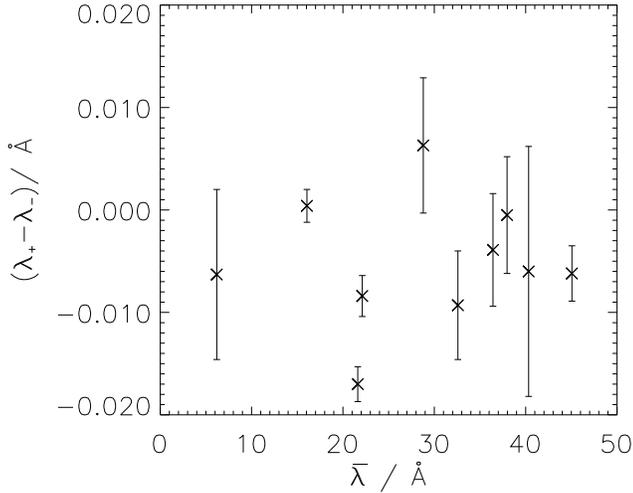}}
 \caption{\label{linlam}Comparison of fit results from left side ($\lambda_-$) and right side ($\lambda_+$) of the spectrum.}
\end{figure}

\section{Observations and data analysis}
\subsection{Observations}
The data described and analyzed in this paper were gathered during the
calibration phase of the {\it Chandra} LETGS. 
Part of the data has already been presented by \cite{brinkman00}.
The individual observation intervals used in our analysis are listed in
Tab.~\ref{dextr}. As is clear
from Tab.~\ref{dextr}, the observations extend over two days for
Procyon and over two months for Capella.
Our analysis refers to the mean properties of Procyon and Capella; the subject 
of possible time variations is not subject of this paper. 
The total on source integration times are 218.54~ksec for Capella and
140.75~ksec for Procyon.
Since the sources are among the strongest coronal X-ray sources
(cf., \cite{huensch98} (1998a,b)), it is also evident that these spectra
belong to the highest quality coronal X-ray spectra
reasonably obtainable with the {\it Chandra} LETGS.

\subsection{Data extraction}

\begin{table}
\caption[ ]{\label{dextr}
List of the data sets with start and stop times and durations used in this paper.}
\begin{flushleft}
\renewcommand{\arraystretch}{1.2}
\begin{tabular}{l r c c}
\hline
Obs.   & time    & \multicolumn{2}{c}{Observation date [UT]} \\
ID.    & [ksec]  & start               & end\\
\hline
\multicolumn{2}{l}{\bf Capella}&&\\
 62435 &  22.34 & 09-06-1999 00:35:40 & 09-06-1999 06:48:01 \\
 01167 &  15.36 & 09-09-1999 13:10:06 & 09-09-1999 17:26:08 \\
 01244 &  12.37 & 09-09-1999 17:42:27 & 09-09-1999 21:08:36 \\
 62410 &  11.33 & 09-09-1999 23:43:57 & 09-10-1999 02:52:48 \\
 01246 &  15.00 & 09-10-1999 03:06:06 & 09-10-1999 07:16:08 \\
 62422 &  11.68 & 09-12-1999 18:26:42 & 09-12-1999 21:41:20 \\
 62423 &  14.80 & 09-12-1999 23:37:44 & 09-13-1999 03:44:28 \\
 01420 &  30.30 & 10-29-1999 22:49:29 & 10-30-1999 07:14:27 \\
 01248 &  85.36 & 11-09-1999 13:42:24 & 11-10-1999 13:25:05 \\
\ total:& 218.54 &&\\
\hline
\multicolumn{2}{l}{\bf Procyon}&&\\
 00063 &  70.39 & 11-06-1999 21:24:31 & 11-07-1999 16:57:38 \\
 01461 &  70.36 & 11-07-1999 17:04:55 & 11-08-1999 12:37:39 \\
\ total:& 140.75 &&\\
\hline
\end{tabular}
\renewcommand{\arraystretch}{1}
\end{flushleft}
\end{table}

All the HRCS datasets analyzed in this paper (see Tab.~\ref{dextr})
were processed using the standard pipeline processing.
The incoming X-rays are diffracted by the grating, dispersing the
different energy photons to different detector positions
along the dispersion direction.
Therefore the spectral information obtained with the LETGS has to be
extracted spatially. 
The pulse heights with which the HRC detector records registered
events contains some very modest energy information, which was,
however, not used.
At each wavelength, the photons have to be integrated in cross-dispersion direction.
Because of the spectrograph's astigmatism, the width of the spectral
trace in cross-dispersion direction is wavelength dependent.
For wavelengths below 75~\AA\ we choose a 3.6\arcsec\ wide extraction
box around the spectral trace which includes almost all of the source signal while
keeping the background level low. For wavelengths greater than 75~\AA\
the extraction box widens in cross dispersion direction to 8.4\arcsec\ at
175~\AA. In addition the photons in four background regions of identical shape
to that of the object extraction region where selected. These regions are
displaced 12\arcsec\ and 24\arcsec\ above and below the source extraction box
in order to check for any spatial variation of the background in
cross-dispersion direction.

\subsection{Symmetry of the grating spectrum}
For the purpose of verifying the symmetry of the dispersed spectra,
we determined 
empirical wavelengths of ten strong emission lines recorded on both sides 
of the middle detector plate (i.e. $\lambda < 50$~\AA). For the example of
Capella we plot the wavelength difference $(\lambda_+-\lambda_-)$
against the mean wavelength $\frac{1}{2}(\lambda_++\lambda_-)$
in Fig.~\ref{linlam}. As can be seen from Fig.~\ref{linlam},
the wavelength values obtained from the right and left side agree to within
less than the instrumental resolution of $\sim$0.06~\AA\
(cf. $\sigma$-values in Tab.~\ref{tab_res}) Both spectra can thus
be added for the analysis in order to increase the SNR of the data.
In the following we will always consider co-added spectra.

\subsection{Method of analysis}
\label{method}

\subsubsection{Treatment of background}
\label{bgfit}
Before determining the flux of any emission line, the spectral
background must be modeled. 
This background consists of essentially two components: instrumental
background and source background, which may consist of continuum radiation
and weak lines.
The instrumental background is also present on those parts of the microchannel
plate not illuminated by X-rays from the target source.
We checked for any variation of the instrumental background perpendicular
to the dispersion direction and found none. We then
modeled this instrumental background $bg$ along the dispersion
direction by a low-order polynomial and
found acceptable fits for almost the whole spectrum.
A far more difficult task is the determination of a reliable source
background $sbg$. The source background is by definition present only
on the extracted spectral trace; it is formed by continuum radiation from
the source, and possibly by weak, unrecognized spectral lines and/or higher
order contamination.\\
For the determination of the fluxes of individual emission lines, the source
background is required only in the vicinity of the line(s) under consideration
and is approximated by a single number. Our numerical procedures
can, however, scope with arbitrary source background models.
In order to estimate this source background, we adopted the following
procedure. We first subtracted the instrumental background from the spectral
trace on the source, and calculated the median of the thus obtained count values
for all bins within the investigated (small) part of the spectrum.
The median is a statistically robust estimate of the background (which we
assume to be flat over the considered part of the
spectrum) as long as more than 50 \% of
all bins belong to the source background, i.e. the spectrum contains not too
many lines.
For each bin $i$ we thus obtain a background value of $sbg+bg_i$
in the studied wavelength range.

\subsubsection{Fit procedure}
The emission line spectra are fitted with a maxi\-mum likelihood technique
similar to that used by \cite{schmitt96a} for their EUVE spectra.
The spectrum is assumed to consist of the background $b$ and of $M$ discrete
emission lines. The treatment of the background $b$ is described in
Sect.~\ref{bgfit}.
Each line $j$ is assumed to be represented by a normalized profile
$g_j(\lambda; \lambda_j,\sigma_j)$, e.g. a Gaussian profile
\begin{equation}
g_j(\lambda;\lambda_j,\sigma_j)=
\frac{1}{\sqrt{2\pi}\sigma_j}{\rm e}^{-\frac{(\lambda-\lambda_j)^2}{2\sigma_j^2}}
\end{equation}
with the dispersion
$\sigma_j$ and the central wavelength $\lambda_j$. The assumption 
of a Gaussian line profile is of course arbitrary. Strong isolated 
emission lines (like O\,{\sc viii} 
Ly$_{\alpha}$) can be fitted quite well with such a model,
and other analytical models can easily be incorporated into our
scheme. 
Let the observed spectrum be given on a grid of $N$ bins with wavelength
values $\lambda_{1}\ldots \lambda_{N}$. The number of expected
counts $c_i$ in the $i^{\rm th}$ bin can then be calculated as
\begin{equation}
\label{ci}
c_i=sbg+bg_i+\sum_{j=1}^{M}a_j g_{i,j}\ ,
\end{equation}
where $a_j$ is the total number of counts of line $j$ and $g_{i,j}$ is the
value of the profile function for line $j$ in bin $i$.
In order to compare the modeled
spectrum $c_i$ with a measured spectrum $n_i$, we assume $n_i$ to be a
Poisson realization of $c_i$. The total probability of the observations
$n_{1}~\ldots~n_{N}$ is then given by
\begin{equation}
P(n_{1}~\ldots~n_{N})=\prod_{i=1}^{N}e^{-c_i}\frac{c_i^{n_i}}{n_i!}
\end{equation}
and the likelihood function ${\cal L}(a_j,\lambda _j,\sigma _j,j=1 \ldots N)$ 
is defined as
\begin{equation}
\label{lP}
{\cal L}= -2 \ \ln P =-2\sum_{i=1}^{N}(-c_i+n_i\ln c_i) +const\ .
\end{equation}
The best-fit values of the parameters $a_j,\ \lambda_j$ and $\sigma_j$ 
$(j=1~$\ldots$~N)$ are 
determined by finding extremal values of $\cal L$ through:
\begin{equation}
\frac{\partial{\, \ln \cal L}}{\partial{\, a_j}}=0\,,\\
\frac{\partial{\, \ln \cal L}}{\partial{\, \lambda_j}}=0\,,\\
\frac{\partial{\, \ln \cal L}}{\partial{\, \sigma_j}}=0\,.
\end{equation}
The physical meaning of the fit parameters $a_j$, $\lambda_j$ and $\sigma_j$
is quite different. $\lambda_j$ and $\sigma_j$ are in principle
fixed by the wavelengths of the considered lines and the instrumental
resolution, complications arise from possible wavelength calibration errors
and line blends. On the other hand, the amplitudes $a_j$, proportional
to the line flux, are the genuine interesting parameters. Analysis of
the likelihood equations shows that
for the amplitude $a_j$ a fixed point equation can be derived for each
line $j$ with $\lambda_j$ and $\sigma_j$ assumed to be given:
\begin{equation}
\label{fixp}
a_{j,_{new}}=\sum^N_{i=1}n_i\frac{a_{j,_{old}}g_{i,j}}{c_{i,_{old}}}\ .
\end{equation}

Eq.~\ref{fixp} can be efficiently solved by iteration.
In order to find optimum values for the wavelengths 
$\lambda_j$ and line-widths $\sigma_j$, we seek minimal values of
$\cal L$ by ordinary minimization procedures. In this process
the wavelengths $\lambda_j$ can vary either freely, or -- if the wavelengths 
of the lines to be fitted are all known -- the wavelength differences between
the individual lines in a multiplet can be kept fixed in order to
account for possible shifts of the overall wavelength scale.
Similarly, the line widths $\sigma_j$ can either vary freely or be fixed,
and in such a way blended lines can be described.
In this fashion
an optimal value for $\cal L$ is obtained, and the parameters
$a_j$, $\lambda _j$ (if fitted) and $\sigma_j$ (if fitted) represent our
best fit measurements of these values.\\
Measurement errors are determined by assuming the likelihood 
curve ${\cal L}(a_j)$
to be parabolic and finding the value of $\Delta a_j$ where
${\cal L}(a_j\pm\Delta a_j)= {\cal L}(a_j) + {\rm d}{\cal L}$ resulting in
$\Delta a_j=\sqrt{2{\rm \Delta}{\cal L}/{\cal L}^{''}_j}$ with ${\cal L}^{''}_j$
being the second derivative of $\cal L$ with respect to $a_j$ known from
Eqs.~\ref{lP} and \ref{ci}.
We choose $\Delta \cal L$\,=\,1 which yields formal 1 $\sigma$ errors.\\
Similarly the errors for $\lambda_j$ and $\sigma_j$ are calculated. We
determine the errors in d$\lambda_j$ and d$\sigma_j$
numerically from ${\cal L}(\lambda_j)$ and ${\cal L}(\sigma_j)$,
respectively with $\Delta \cal L$\,=\,1, i.e. errors being given within 68.3\%.
Thus formally we treat all other parameters as ``uninteresting''.

\subsection{Extracted spectra and measured line ratios}
\begin{table*}
\caption[ ]{\label{tab_res}Measured line ratios for Capella and Procyon}
\begin{flushleft}
\renewcommand{\arraystretch}{1.2}
\begin{tabular}{r|c|c|c|c|c|c}
{\bf Capella}&$\lambda$\ [\AA]&$\sigma$\ [\AA]&A\ [cts]&sbg\ [cts/\AA]&$R_{\rm obs}=$f/i&$G_{\rm obs}=\frac{\rm i+f}{\rm r}$\\
\hline
\hline
O\,{\sc vii}&&&&&&\\
r&21.62 $\pm$ 0.005&&3071.2 $\pm$ 56.0&&&\\
i&21.82 $\pm$ 0.007&0.027&$544.8 \pm 31.4$&2997&3.92$\pm$0.24&0.87$\pm$0.03\\
f&22.12 $\pm$ 0.001&&2135.2 $\pm$ 51.1&&&\\
\hline
N\,{\sc vi}&&&&&&\\
r&28.79 $\pm$ 0.003&&491.2 $\pm$ 31.49&&&\\
i&29.1 $\pm$ 0.004&0.03&228.2 $\pm$ 26.5&3265&1.68$\pm$0.23&1.25$\pm$0.14\\
f&29.54 $\pm$ 0.003&&384.5 $\pm$ 29.4&&&\\
\hline
C\,{\sc v}&&&&&&\\
r&40.28 $\pm$ 0.002&0.026&440.7 $\pm$ 26.9&&&\\
i&40.72 $\pm$ 0.008&0.026&101.3 $\pm$ 18.24&1200&1.58$\pm$0.36&0.59$\pm$0.1\\
f&41.5 $\pm$ 0.005&0.029&160.2 $\pm$ 22.37&&&\\
{\bf Procyon}&&&&&&\\
\hline
\hline
O\,{\sc vii}&&&&&&\\
r&21.6 $\pm$ 0.006&&731.6 $\pm$ 28.7&&&\\
i&21.8 $\pm$ 0.004&0.027&203 $\pm$ 16.8&0&3.21$\pm$0.3&1.17$\pm$0.08\\
f&22.1 $\pm$ 0.003&&652.4 $\pm$ 27.3&&&\\
\hline
N\,{\sc vi}&&&&&&\\
r&28.8 $\pm$ 0.003&&200.2 $\pm$ 16.8&&&\\
i&29.1 $\pm$ 0.006&0.03&77.4 $\pm$ 12.3&0&1.26$\pm$0.26&0.87$\pm$0.14\\
f&29.55 $\pm$ 0.005&&97.1 $\pm$ 13.2&&&\\
\hline
C\,{\sc v}&&&&&&\\
r&40.28 $\pm$ 0.003&0.03&203.8 $\pm$ 17.0&&&\\
i&40.75 $\pm$ 0.004&0.028&123.1 $\pm$ 14.2&0&0.51$\pm$0.12&0.92$\pm$0.16\\
f&41.48 $\pm$ 0.01&0.046&63.4 $\pm$ 13.5&&&\\
\hline
\end{tabular}
\renewcommand{\arraystretch}{1}
\end{flushleft}
\end{table*}

With the procedure described in Sect.~\ref{method} we analyzed the spectra of
the He-like
triplets of O\,{\sc vii}, N\,{\sc vi}, and C\,{\sc v} for the two stars Capella and Procyon.
In Figs.~\ref{proc_spectrum} and \ref{cap_spectrum} the measured spectra are
plotted with bold line-dotted lines and the best-fit model curve is indicated
by a thin solid line.
The total background, i.e. the instrumental background and the assumed
source continuum background, is shown with a dotted line.
The derived best-fit parameters are given in Tab.~\ref{tab_res},
where we list, for both Capella and Procyon, the derived
empirical wavelengths, the line widths and the line strengths (in counts).
For reference purposes we also list the source background values used.
Extrapolating the instrumental background onto the spectral trace shows
that Procyon's source background must be quite small as expected for a
low coronal temperature X-ray source. It will therefore be neglected for the
purpose of line flux modeling. In Tab.~\ref{tab_res} we also list the ratios
between forbidden and intercombination line, and that of the
sum of intercombination and forbidden line to the resonance line, which
are needed for subsequent analysis. Since no significant
rotational or orbital line broadening is expected given the even high spectral
resolution of our {\it Chandra} data, the line-width $\sigma$ was kept fixed
for all models assuming that this value described the instrumental resolution.
Starting values for the wavelengths were taken from \cite{mewe85}.\\
The O\,{\sc vii} triplet has the best signal to noise ratio in both stars and is
unaffected by any significant contamination from higher orders
as can be seen from
Fig.~\ref{proc_spectrum} and Fig.~\ref{cap_spectrum} in the top panel.
As far as the N\,{\sc vi} triplet is concerned, it
is not as isolated as the O\,{\sc vii} triplet and requires further
analysis. In both Procyon (Fig.~\ref{proc_spectrum}) and Capella
(Fig.~\ref{cap_spectrum}) additional lines appear.
At 28.44~\AA\ a line attributed to C\,{\sc vi} is evident for both Capella and
Procyon. In order to minimize any possible
cross talk, this line was included in the fit.
The line at 30~\AA\ seen in the Capella spectrum
is interpreted as the second order of the strong Fe\,{\sc xvii} line at 15.013~\AA;
Fe\,{\sc xvii} is strong in Capella, but essentially absent in Procyon. 
A somewhat strange feature can be seen only in the N\,{\sc vi}
spectrum of Procyon at about
29.3~\AA. This is not a line but an instrumental effect which is present
only on the negative side of the spectrum.

\subsubsection{Analysis of the C\,{\sc v} triplet}
\begin{table}
\caption[ ]{\label{lo_list} Fit result in the 13\,--\,14\,\AA\ range for Capella. The
assumed source background is 5792 cts/\AA}
\begin{flushleft}
\renewcommand{\arraystretch}{1.2}
\begin{tabular}{l|c|c|c}
Line&$\lambda$\ [\AA]&$\sigma$\ [\AA]&A\ [cts]\\
\hline
Ne\,{\sc ix}&$13.46\pm 0.003$&0.025&$4373.9\pm 82.92$\\
Fe\,{\sc xix}&$13.53\pm 0.001$&0.025&$4683.4\pm 84.52$\\
Ne\,{\sc ix}&$13.71\pm 0.004$&0.033&$3019.9\pm 74.50$\\
Fe\,{\sc xix}&$13.8\pm 0.02$&0.02&$496.6\pm 62.06$\\
Fe\,{\sc xvii}&$13.83\pm 0.008$&0.03&$3337.6\pm 77.94$\\
\end{tabular}
\renewcommand{\arraystretch}{1}
\end{flushleft}
\end{table}

\begin{figure}
 \resizebox{\hsize}{!}{\includegraphics{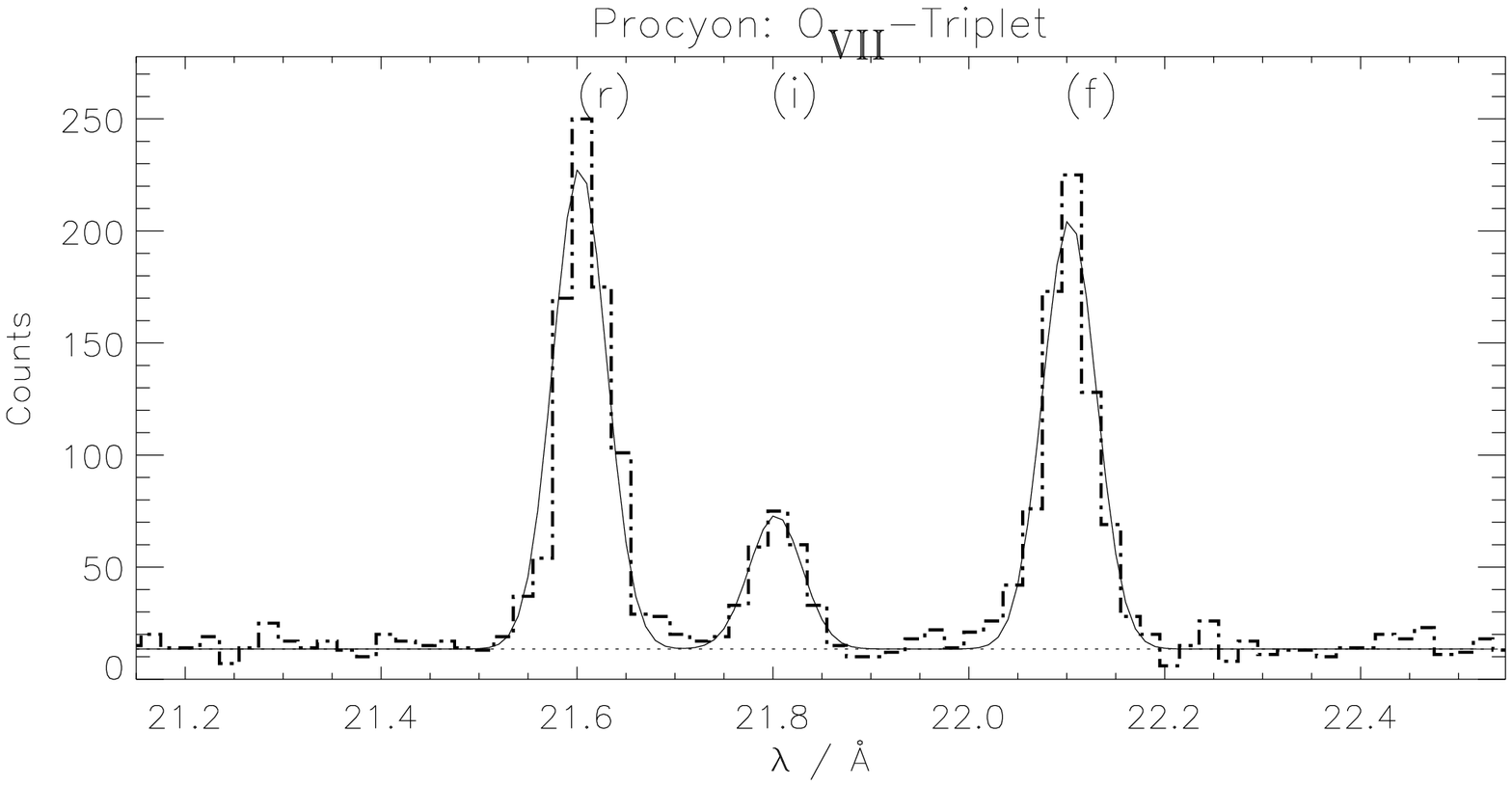}}
 \resizebox{\hsize}{!}{\includegraphics{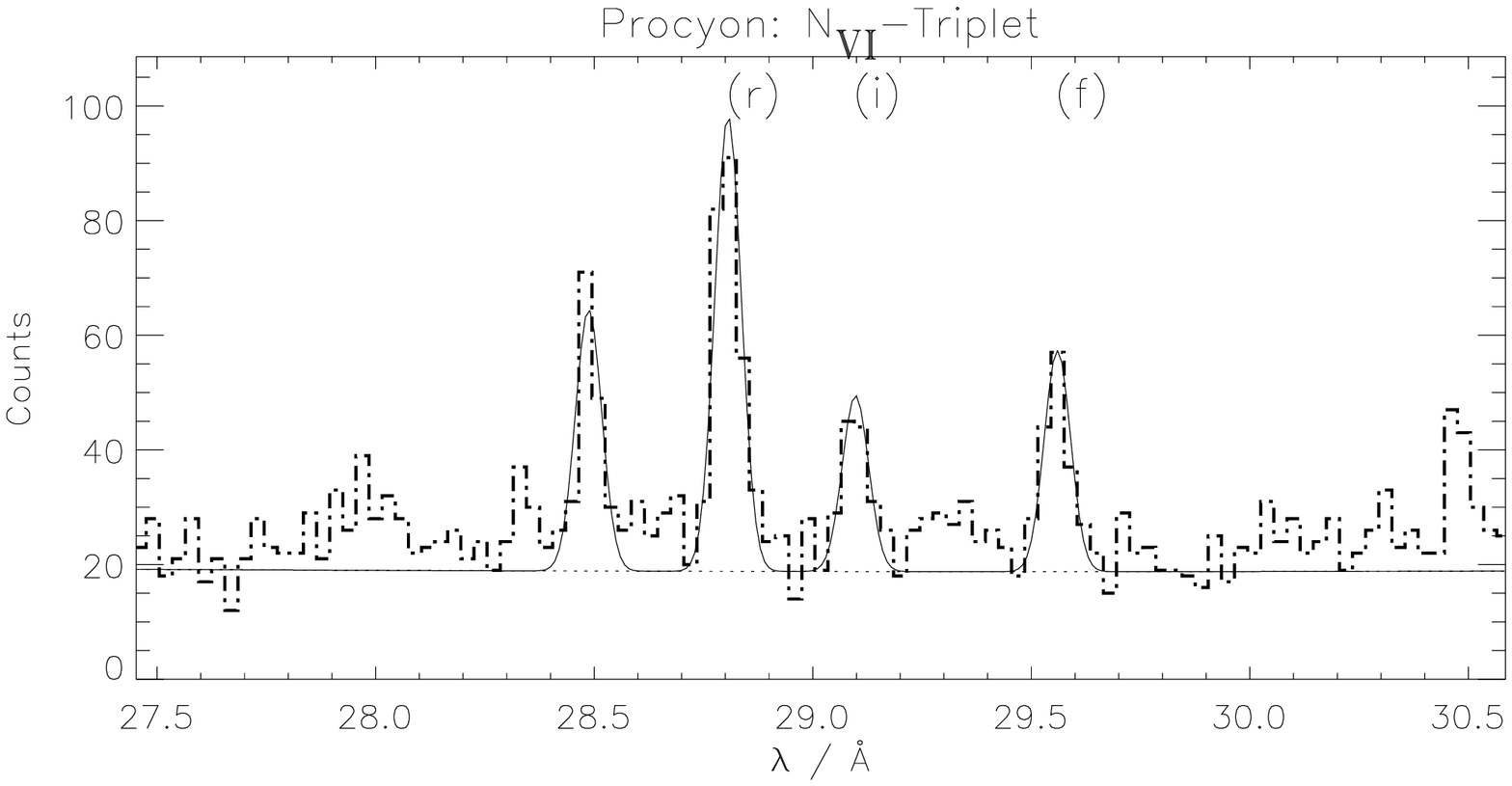}}
 \resizebox{\hsize}{!}{\includegraphics{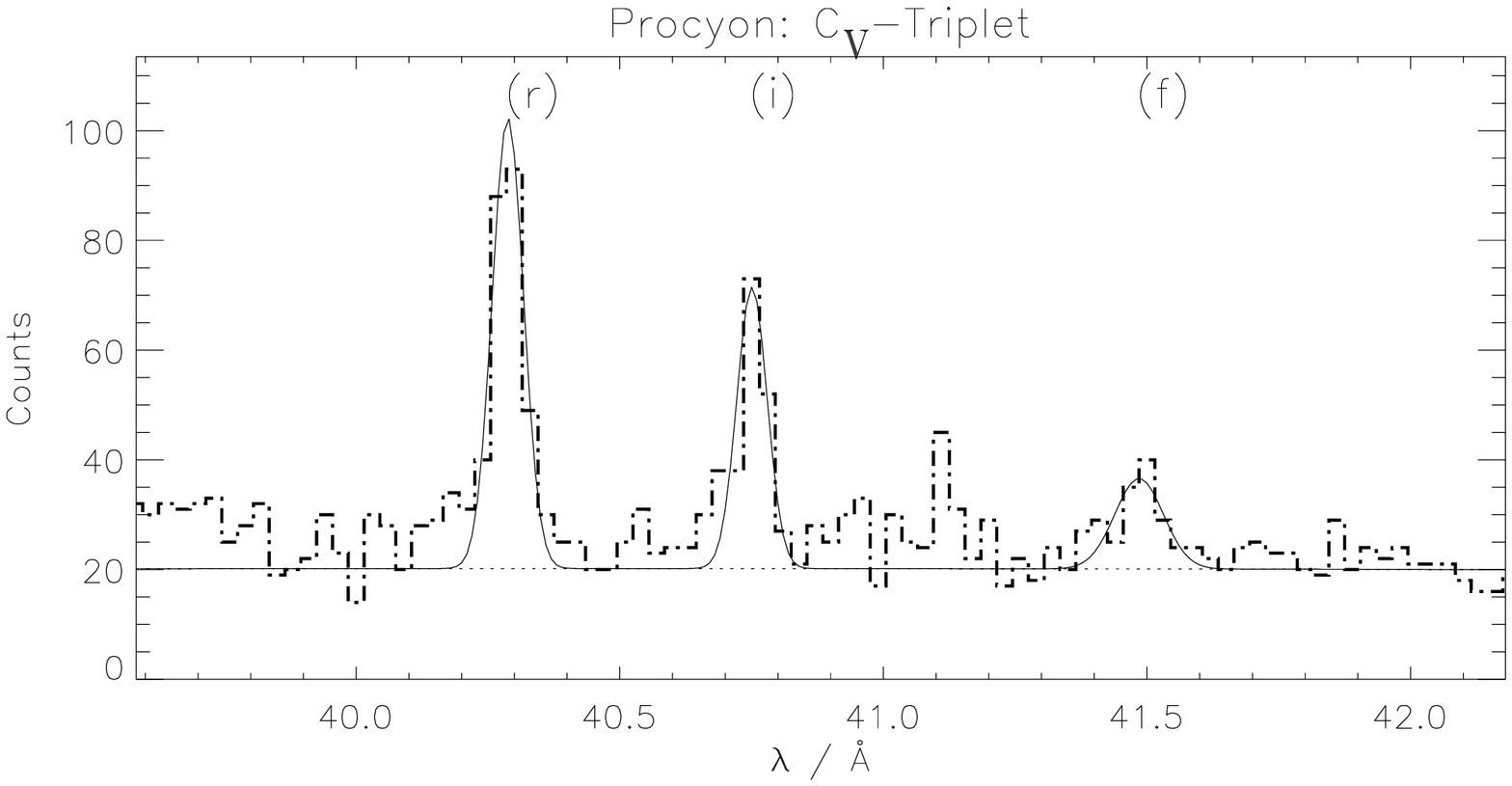}}
 \caption{Spectrum (line-dotted) and fitted (solid) curve for the O\,{\sc vii}, N\,{\sc vi},
and C\,{\sc v} triplet for Procyon. The dotted line represents the total background.
The binsize is 0.02~\AA\ for O\,{\sc vii} and 0.03~\AA\ for C\,{\sc v} and N\,{\sc vi}.}
 \label{proc_spectrum}
\end{figure}

\begin{figure}
 \resizebox{\hsize}{!}{\includegraphics{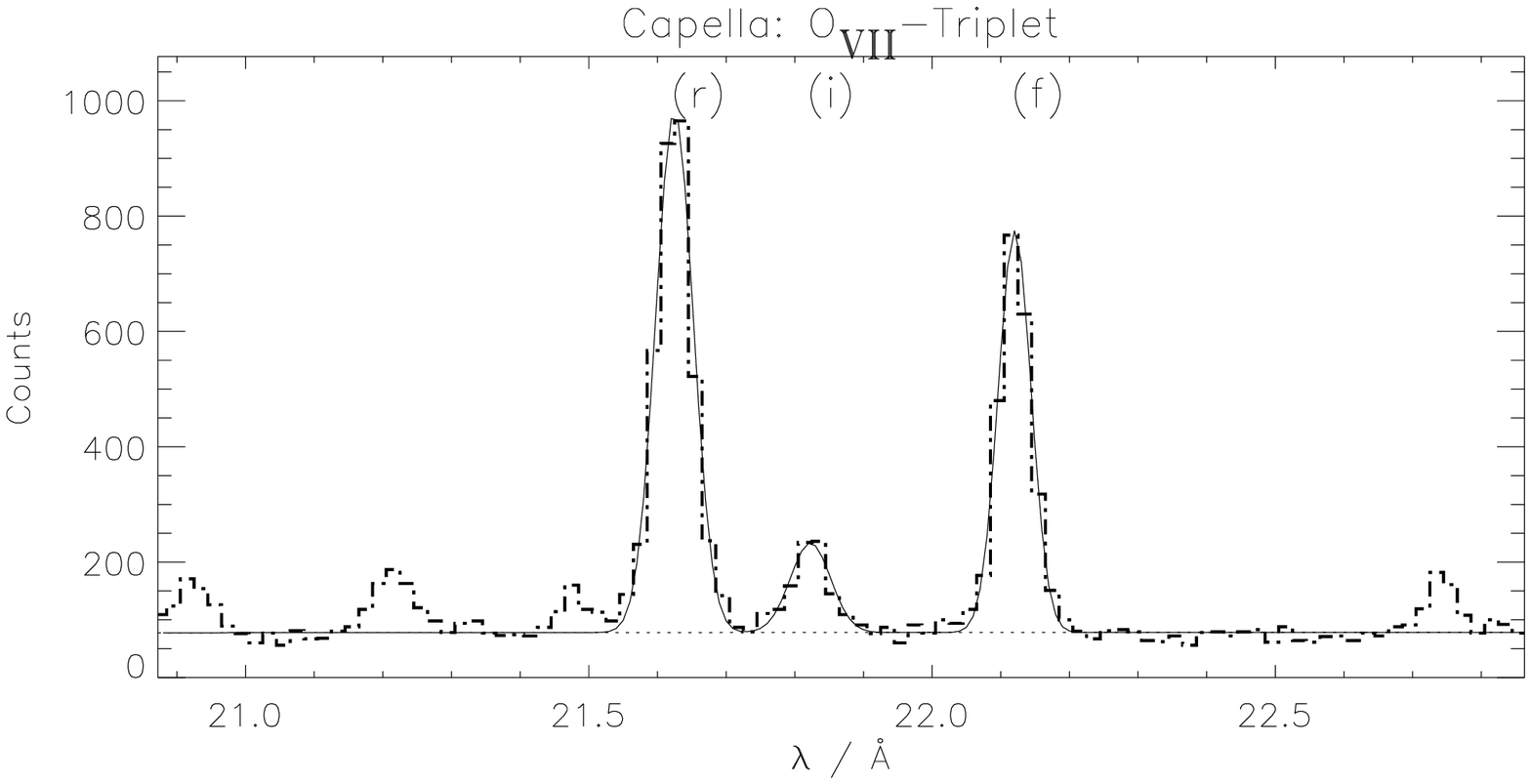}}
 \resizebox{\hsize}{!}{\includegraphics{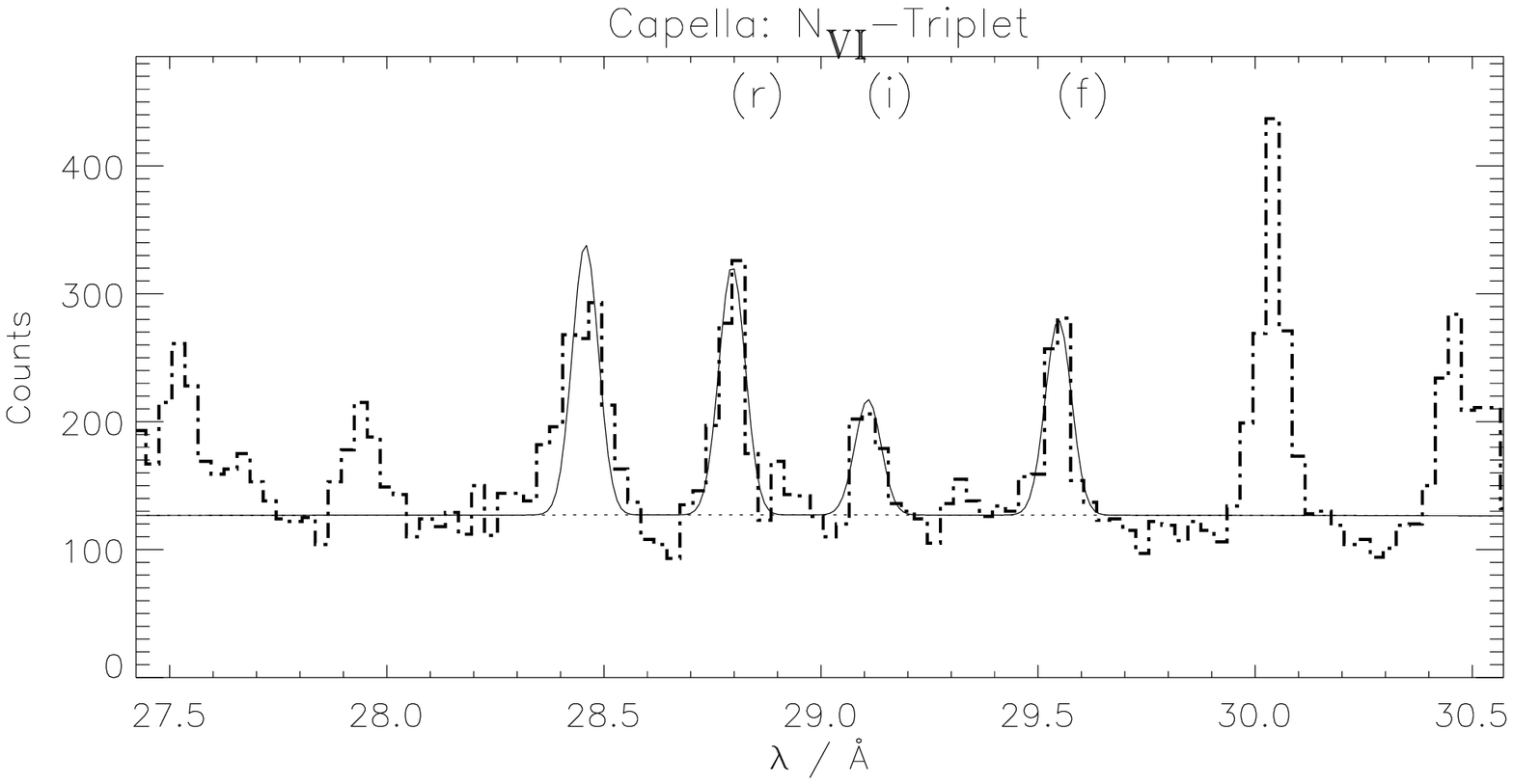}}
 \resizebox{\hsize}{!}{\includegraphics{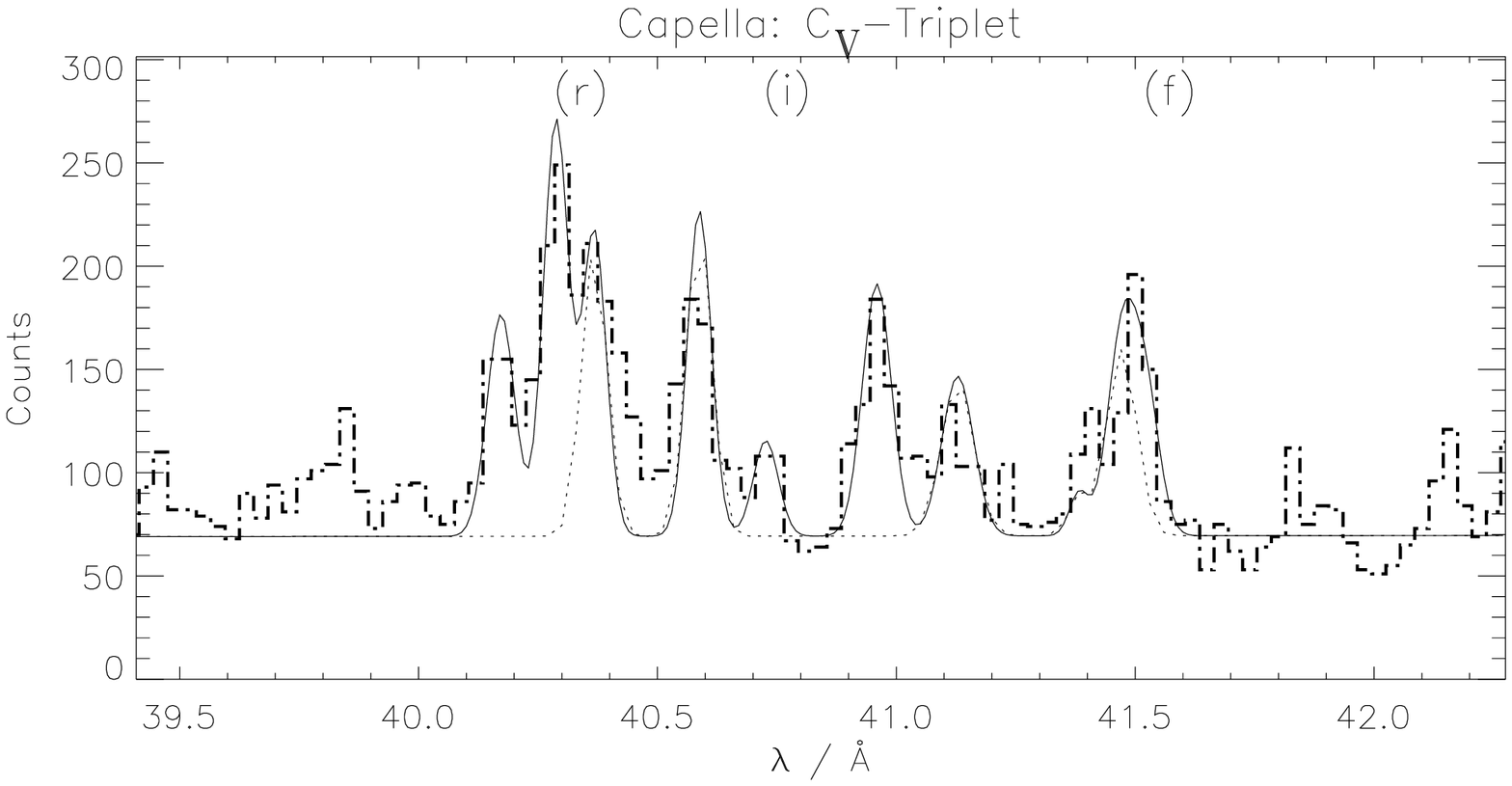}}
 \caption{Spectrum (line-dotted) and fitted (solid) curve for the O\,{\sc vii}, N\,{\sc vi},
and C\,{\sc v} triplet for Capella. The dotted line represents the total background.
The binsize is 0.02~\AA\ for O\,{\sc vii} and 0.03~\AA\ for C\,{\sc v} and N\,{\sc vi}.}
 \label{cap_spectrum}
\end{figure}

\begin{figure}
 \resizebox{\hsize}{!}{\includegraphics{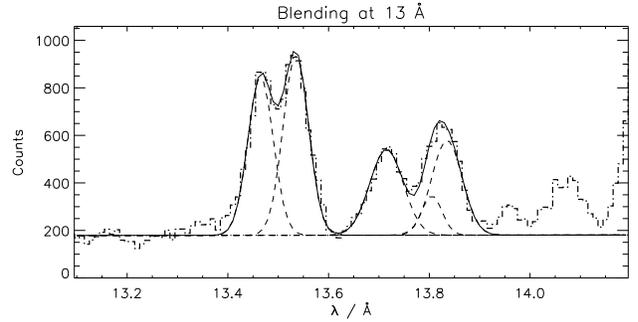}}
 \caption{First order spectrum responsible for third order contamination
of the C\,{\sc v} triplet of Capella; shown are the data (dash-dotted histogram) as well as the best fit (solid line). Note that the feature near 13.8 \AA\ is actually
a line blend (consisting of Fe\,{\sc xix} and Fe\,{\sc xvii} contributions) resolved in third order.}
 \label{lo_1}
\end{figure}

Contamination with higher order lines makes the
analysis of the C\,{\sc v} triplet for Capella particularly difficult (cf.,
Fig.~\ref{cap_spectrum}). This is clear since the Capella spectrum contains
strong emission in the band between 13\,--\,14~\AA, which appears in third
order in the 39\,--\,42 \AA\ band under consideration for the C\,{\sc v} triplet.
Specifically, the interfering lines are Ne\,{\sc ix} (13.44~\AA\ $\times~3$
= 40.32~\AA), Fe\,{\sc xix} (13.52~\AA\ $\times~3$ = 40.56~\AA), Ne\,{\sc ix} (13.7~\AA\
$\times~3$ = 41.1~\AA), Fe\,{\sc xix} (13.795~\AA\ $\times~3$ = 41.385~\AA)
and Fe\,{\sc xvii} (13.824~\AA\ $\times~3$ = 41.472~\AA). We modeled the
contamination from these five lines by first determining their
first order contributions, transforming the fit results to third order.
In the transformation the first order intensities are reduced by a factor
of 14.1. This reduction factor is obtained by comparison of first and third
order of the isolated 15.013~\AA\ line (Fe\,{\sc xvii}). Since the contaminating
photons are rather energetic, we expect essentially the same reduction factor
in the range 13\,--\,14~\AA. The first order fit result is plotted in
Fig.~\ref{lo_1} and the resulting fit values are listed in
Tab.~\ref{lo_list}. Technically
the higher order contamination was treated as an additional nonconstant
contribution to the instrumental background. In addition to the
C\,{\sc v} triplet, another line, Si\,{\sc xii} at 40.91~\AA, appearing strong in 
Capella but weaker in Procyon, had to be modeled.
The source background was estimated by requiring the lowest count bins to
be adequately modeled. The median function could not
be applied since there are too many lines in the considered wavelength range.
The result of this
modeling exercise is shown in Fig.~\ref{cap_spectrum} in the bottom panel,
where the dotted line, representing the background, is not constant, 
but heavily influenced by higher order lines. We emphasise that
the errors listed in Tab.~\ref{tab_res} do not include errors from the fits
in the 13~\AA\ band. We point out in particular
that the forbidden C\,{\sc v} line lies on top of the third order Fe\,{\sc xvii}
13.824~\AA\ line, so that any derivation of the forbidden line flux
does require an appropriate modeling of the third order contamination. This
is especially difficult because the Fe\,{\sc xvii} 13.824~\AA\ line is blended
with the Fe\,{\sc xix} 13.795~\AA\ line in first order, while
in third order this blend is resolved.\\
The strong emission in the 13~\AA\ regime is mostly due to iron in 
excitation stages Fe\,{\sc xvii} and Fe\,{\sc xix}.
Since there is no significant emission from iron in these high excitation stages
in the spectrum of Procyon, there is no blending with
third order lines in this case. From Fig.~\ref{proc_spectrum} it can be seen
that the modeling is straightforward for Procyon.\\

%#######################################################
%               Results and Interpretation
%#######################################################

\section{Results and interpretation}

Inspection of our Chandra spectra of Capella and Procyon shows the O\,{\sc vii}
and N\,{\sc vi} forbidden lines to be present in both cases. The C\,{\sc v} forbidden line
is very weak but present for Procyon, it is also present in Capella despite
being significantly contaminated by third order radiation. The observed
f/i-ratios for O\,{\sc vii} are very close to the expected low-density limit $R_{\rm 0}$
for both Procyon and Capella, while the measured f/i-ratios for N\,{\sc vi} and
C\,{\sc v} are definitely below the respective values of $R_{\rm 0}$ in both cases.
Before giving a quantitative interpretation of the observed line ratios we must
first consider the formation temperature of the studied He-like lines.
Especially for the C\,{\sc v} triplet the stellar radiation field can contribute
significantly to the depopulation of the atomic level from which the forbidden 
line originates. Low forbidden line intensities can thus also indicate
large radiation fields and not necessarily high densities.

\subsection{Formation temperature of lines}

\begin{table}
\caption{\label{ftemp} Temperature diagnostics from ratio of Ly$_{\alpha}$/r and
from the ratio $G$\,=\,(i+f)/r. Values for r and $G$
are taken from Tab.~\ref{tab_res} (the $G$ values listed are corrected for
effective areas; cf., Tab.~\ref{eff_ar}). For comparison the peak line formation
temperature $T_{\rm m}$ is also given. $T(G)$ is calculated according to
\cite{mewe00}.}
\begin{flushleft}
\renewcommand{\arraystretch}{1.2}
\begin{tabular}{l c r r}
\hline
Ion & $\lambda$ & Procyon & Capella\\
    &   [\AA]   &  [cts]  &[cts]  \\ 
\hline
C\,{\sc vi} (cts)     & 33.75 & 697.5  $\pm$ 28.1 & 2151.1  $\pm$ 51.3\\
C\,{\sc v}  (cts)     & 40.27 & 203.8  $\pm$ 17.0 & 440.7   $\pm$ 26.9\\
\ C\,{\sc vi}/C\,{\sc v}     &       & 3.42   $\pm$ 0.32 & 4.88    $\pm$  0.41\\
$T$(H-He)/MK   &       & 1.14   $\pm$ 0.03 & 1.27    $\pm$ 0.03\\
$G_{\rm obs}$  &       & 1.27   $\pm$ 0.21 & 0.81    $\pm$ 0.14\\
$T(G)$/MK      &       & 0.32   $\pm$ 0.15 & 0.98    $\pm$ 0.43\\
$T_{\rm m}$/MK &       & 1.0               & 1.0\\
\hline
N\,{\sc vii} (cts)    & 24.8  & 206.9  $\pm$ 16.95 & 2280.3 $\pm$ 53.6\\
N\,{\sc vi}  (cts)    & 28.79 & 200.2  $\pm$ 16.8 & 491.2  $\pm$ 31.49\\
\ N\,{\sc vii}/N\,{\sc vi}   &       & 1.03   $\pm$ 0.12  & 4.64   $\pm$ 0.41 \\
$T$(H-He)/MK   &       &$1.5\pm 0.05$&$2.5\pm 0.08$\\
$G_{\rm obs}$  &       & 0.93   $\pm$ 0.16 & 1.33    $\pm$ 0.15\\
$T(G)$/MK      &       & 1.25   $\pm$ 0.60 & 0.47    $\pm$ 0.17\\
$T_{\rm m}$/MK &       & 1.4              & 1.4\\
\hline
O\,{\sc viii} (cts)   & 18.97 & 673.2 $\pm$ 27.6 & 14676.8 $\pm$ 124.3\\
O\,{\sc vii}  (cts)   & 21.6  & 731.6  $\pm$ 28.7 & 3071.2 $\pm$ 56\\
\ O\,{\sc viii}/O\,{\sc vii} &       & 0.92   $\pm$ 0.05  & 4.78   $\pm$ 0.13 \\
$T$(H-He)/MK   &       & $2.13\pm 0.03$&$3.4\pm 0.03$\\
$G_{\rm obs}$  &       & 1.21   $\pm$ 0.08 & 0.90    $\pm$ 0.03\\
$T(G)$/MK      &       & 1.0    $\pm$ 0.16 & 2.01    $\pm$ 0.16\\
$T_{\rm m}$/MK &       & 2.0               & 2.0\\
\hline
\end{tabular}
\renewcommand{\arraystretch}{1}
\end{flushleft}
\end{table}

For the purpose of density diagnostics it is customary to assume
that all of the emission is produced at a single temperature $T_{\rm m}$ 
(cf., Tab.~\ref{ionvals}), which corresponds
to the peak of the contribution function of the considered line, the so-called
formation temperature. It should
be kept in mind, however, that He-like ions are present over a relatively
broad temperature range, and therefore this assumption might be poor 
if steep emission measure gradients are present. The measured value of
$G$ is also a temperature diagnostics, although optical depth effects in the
resonance line may contribute (cf., \cite{acton78});
the relevance of optical depth effects to our results is analyzed in
section~\ref{optdepth}.
In Tab.~\ref{ftemp} we show the temperatures $T$(G) determined from converting
the observed $G_{\rm obs}$ values into temperatures according to \cite{mewe00}
using an electron density of $n_e=5\cdot 10^9$~cm$^{-3}$.
We also determined temperatures by studying the line ratio between the observed
Ly$_{\alpha}$ lines of O\,{\sc viii}, N\,{\sc vii}, and C\,{\sc vi} and the r-lines of O\,{\sc vii}, N\,{\sc vi}, and C\,{\sc v}
respectively by assuming isothermal plasma emission. We assume plasma
emissivities as calculated in the codes MEKAL (\cite{mewe85}, \cite{mewe95})
and SPEX (\cite{kaastra96}). The thus obtained fluxes are multiplied with
effective areas (cf., Tab.~\ref{eff_ar}) before comparison with the measured
ratios. The results are listed in Tab.~\ref{ftemp} as $T$(H-He).
\begin{table}
\caption[ ]{\label{eff_ar}Effective areas of the detector at the designated
wavelengths. The values are taken from In-Flight Calibration by Deron Pease
(9 March 2000).}
\begin{flushleft}
\renewcommand{\arraystretch}{1.2}
\begin{tabular}{r r r r r}
\hline
element & $\lambda$ / \AA &  \multicolumn{3}{c}{A$_{\rm eff}$ / cm$^2$} \\
        &                 &  $\lambda_+$    &  $\lambda_-$   &  total\\
C\,{\sc vi}     & 33.76           &  5.09           &  5.09          & 10.18\\
C\,{\sc v} (r)  & 40.28           &  2.34           &  2.34          & 4.68\\
C\,{\sc v} (i)  & 40.74           &  1.65           &  1.65          & 3.30\\
C\,{\sc v} (f)  & 41.5            &  1.76           &  1.76          & 3.52\\
\hline
N\,{\sc vii}    & 24.8            &  7.76           &  7.75          & 15.51\\
N\,{\sc vi} (r) & 28.8            &  6.34           &  6.33          & 12.67\\
N\,{\sc vi} (i) & 29.1            &  6.16           &  6.15          & 12.31\\
N\,{\sc vi} (f) & 29.54           &  5.80           &  5.79          & 11.59\\ 
\hline
O\,{\sc viii}   & 18.98           &  10.08          &  10.70         & 20.78\\
O\,{\sc vii} (r)& 21.62           &  6.47           &  7.05          & 13.52\\
O\,{\sc vii} (i)& 21.82           &  6.36           &  6.94          & 13.30\\
O\,{\sc vii} (f)& 22.12           &  6.22           &  6.81	     & 13.03\\
\hline
\end{tabular}\\
\renewcommand{\arraystretch}{1}
\end{flushleft}
\end{table}

As can be seen from Tab.~\ref{ftemp}, the temperatures $T$(G) and
$T$(H-He) do not agree. This is not surprising given the fact that we are
likely dealing with a temperature distribution. For Procyon the
temperatures $T$(H-He) agree quite well with $T_m$, while
$T$(G) agrees with $T_m$ only for N\,{\sc vi}. For Capella the respective
temperatures $T$(H-He) always exceed those found for Procyon, while
the $T$(G) temperature derived from N\,{\sc vi} is below that found for Procyon.
We tentatively conclude that the observed N\,{\sc vi} and O\,{\sc vii}
emission in Capella has significant contributions from plasma at
temperatures away from the peak in the line emissivity curve, while for
Procyon the emission appears to come from rather close to the
line emissivity peak.

\subsection{Influence of the stellar radiation field}
\label{rad_field}

The observed C\,{\sc v} line ratios are particularly interesting. While
the observed value for Capella agrees well with the solar observations
(cf., \cite{aus66}, \cite{freeman70}), the observed value for Procyon is
rather small. The important point to keep in mind in this context 
is that Procyon
and Capella are stars with different properties compared to the Sun. Procyon 
is of spectral type F5V-IV (cf., Tab.~\ref{star_prop}) with an effective
temperature of 6500~K, Capella is a spectroscopic binary,
the components of which are of spectral type G1 and G8; occasionally
the brighter component is also classified as F9.
Strictly speaking, what really
matters is the effective temperature of the radiation field at the
wavelength corresponding
to the energy difference between forbidden and intercombination line
levels, i.e., 2272 \AA\ for C\,{\sc v} (cf., Tab.~\ref{rad_temp}) for the
transition 2~$^3P\rightarrow 2\ ^3S$. We investigated the stellar
surface radiation fluxes from measurements obtained with the International 
Ultraviolet Explorer satellite (IUE). We first determined continuum
fluxes from archival IUE data, and converted these fluxes into
intensities using the expression
\begin{equation}
I_\lambda=F_\lambda \frac {d^2} {R^2} \frac {1} {2\pi(1-\frac{\epsilon}{2})},
\end{equation}
where $\epsilon$ is the (linear) limb darkening coefficient.
It is determined
from \cite{diaz95} using log~$g$ and $T_{\rm eff}$ listed in 
Tab.~\ref{star_prop}; $d$ and $R$ are distance and radius of the sample
stars as listed in Tab.~\ref{star_prop}.
For the case of the binary system Capella we assumed the worst case and
attributed all emission to the Ab component, i.e., the star with the higher 
radiation temperature. The other component Aa (G8/K0~III) is about equally
bright in the corona as G1~III (cf., \cite{linsky98}),
but much weaker in the chromosphere and transition region.
From the thus obtained value for the intensity a radiation temperature
$T_{\rm rad}$ can be inferred from the appropriate Planck curve, expressed as
\begin{equation}
u_\nu = W \frac{8\pi h \nu^3}{c^3} \frac{1}{{{\rm exp}\Bigl({{h \nu}\over {kT_{\rm rad}}}\Bigr) - 1}},
\end{equation}
where $W$ is the dilution factor of the radiation field (we take $W={1\over 2}$ close to the stellar
surface).
The results derived from these data are listed in Tab.~\ref{rad_temp}.\\

\begin{table}
\caption[ ]{\label{rad_temp}Investigation of the influence of the
stellar radiation field. Measured fluxes from the IUE satellite $F_{\lambda}$
are converted to intensity $I_{\lambda}$ taking into account limb darkening effects
using $\epsilon$ from Tab.~\ref{star_prop}.}
\begin{flushleft}
\renewcommand{\arraystretch}{1.2}
\begin{tabular}{r r r r}
\hline
&C\,{\sc v}&N\,{\sc vi}&O\,{\sc vii}\\
\hline
$\lambda_{\rm f\rightarrow i}$/\AA&2272&1900&1630\\
$I_{\rm pot}$/eV&392.1&552.1&739.3\\
%-----
\multicolumn{4}{c}{\bf Capella}\\
$F_{\lambda}/(10^{-11}\frac{\rm ergs}{\rm cm^{2}\,s\,
\textrm{\tiny \AA}}$)&$2\pm0.5$&$1.2\pm0.5$&$0.25\pm0.05$\\
$I_{\lambda}/(10^4\frac{\rm ergs}{\rm cm^{2}\,s\,
\textrm{\tiny \AA}\,strd}$)&$1.98\pm0.5$&$1.19\pm0.3$&$0.25\pm0.5$\\
$T_{\rm rad}$/K&$4585\pm100$&$4976\pm150$&$5029\pm50$\\
$\phi/\phi_{\rm c}$  & $2.54\pm0.86$ & $0.2\pm0.1$ & 0.003\\
&&&$\pm0.0005$\\
\hline
%-----
\multicolumn{4}{c}{\bf Procyon}\\
$F_{\lambda}/(10^{-11}\frac{\rm ergs}{\rm cm^{2}\,s\,\textrm{\tiny \AA}}$)&$15\pm4$&$7\pm3$&$0.6\pm0.3$\\
$I_{\lambda}/(10^4\frac{\rm ergs}{\rm cm^{2}\,s\,{\textrm{\tiny \AA}}\,strd}$)&$21\pm5$&$9.8\pm4$&$0.84\pm0.4$\\
$T_{\rm rad}$/K&$5532\pm150$&$5778\pm200$&$5406\pm300$\\
$\phi/\phi_{\rm c}$  & $26.67\pm9.3$ & $1.58\pm0.84$ & $0.01\pm0.015$\\
$\phi_{\rm c}$/(s$^{-1}$)       & 34.6           & 148           & 717 \\
\hline
\end{tabular}
\renewcommand{\arraystretch}{1}
\end{flushleft}
\end{table}

In order to compute the dependence of $\phi/\phi_{\rm c}$, required 
in Eq.~\ref{R_Ne}, on $T_{\rm rad}$ we use the
calculations by \cite{blume72}, who derive the expression
\begin{equation}
\label{phi_phic}
\frac{\phi}{\phi_{\rm c}}=\frac{3(1+F)c^3}{8\pi h \nu^3}
\frac{A(2 ^3P\rightarrow 2 ^3S)}{A(2 ^3S_1 \rightarrow\ 1 ^1S_0)}u_\nu
\end{equation}
with $\phi_{\rm c} = {A(2 ^3S_1 \rightarrow\ 1 ^1S_0)}/(1+F)$, transition 
probabilities 
$A(i\rightarrow j)$, and $u_\nu$ the spectral energy density
at the appropriate $2 ^3P_1\rightarrow\ 2 ^3S_1$ frequency $\nu$. 
The factor 3 is the ratio of the statistical weights of the levels $2 ^3P$ and
$2 ^3S$. $F$ is approximated by \cite{blume72} through
\begin{equation}
F(\xi)=\frac{3\xi H(\xi) exp(\xi/4)+1.2 \xi +2H(\xi)+2}
{3\xi exp(\xi/4)+0.6\xi H(\xi)+H(\xi)+1},
\end{equation}
where
\begin{equation}
H(\xi)=\frac{C(1 ^1S\rightarrow\ 2 ^3S)}{C(1 ^1S\rightarrow\ 2 ^3P)},
\end{equation}
with $\xi=I/kT_{\rm rad}$ and the collisional excitation rate coefficients
$C(i\rightarrow j)$.
The ionization potentials $I_{\rm pot}$ for the
ions C\,{\sc v}, N\,{\sc vi}, and O\,{\sc vii} are taken from \cite{pradhan81b}.
The collisional excitation rate coefficients $C(i\rightarrow j)$ were taken
from \cite{pradhan81a}
for C\,{\sc v} and O\,{\sc vii} while the rate coefficients for N\,{\sc vi}
were obtained by interpolation from these values. \\
From Eq.~\ref{phi_phic} we determine the values used in Eq.~\ref{R_Ne}
for the calculation of the theoretical curves in Fig.~\ref{dens}.
They are also listed in Tab.~\ref{rad_temp}.
For comparison, we also give in Tab.~\ref{rad_temp} the values derived from
recent calculations by \cite{mewe00} considering a 
multi-level model previously used by \cite{porquet00},
and taking into account the effect of temperature on $G$ and $R(n_e)$ 
(\cite{mewe00}, see also \cite{porquet00}).

\subsection{Analysis and interpretation}

\begin{figure}
 \resizebox{\hsize}{!}{\includegraphics{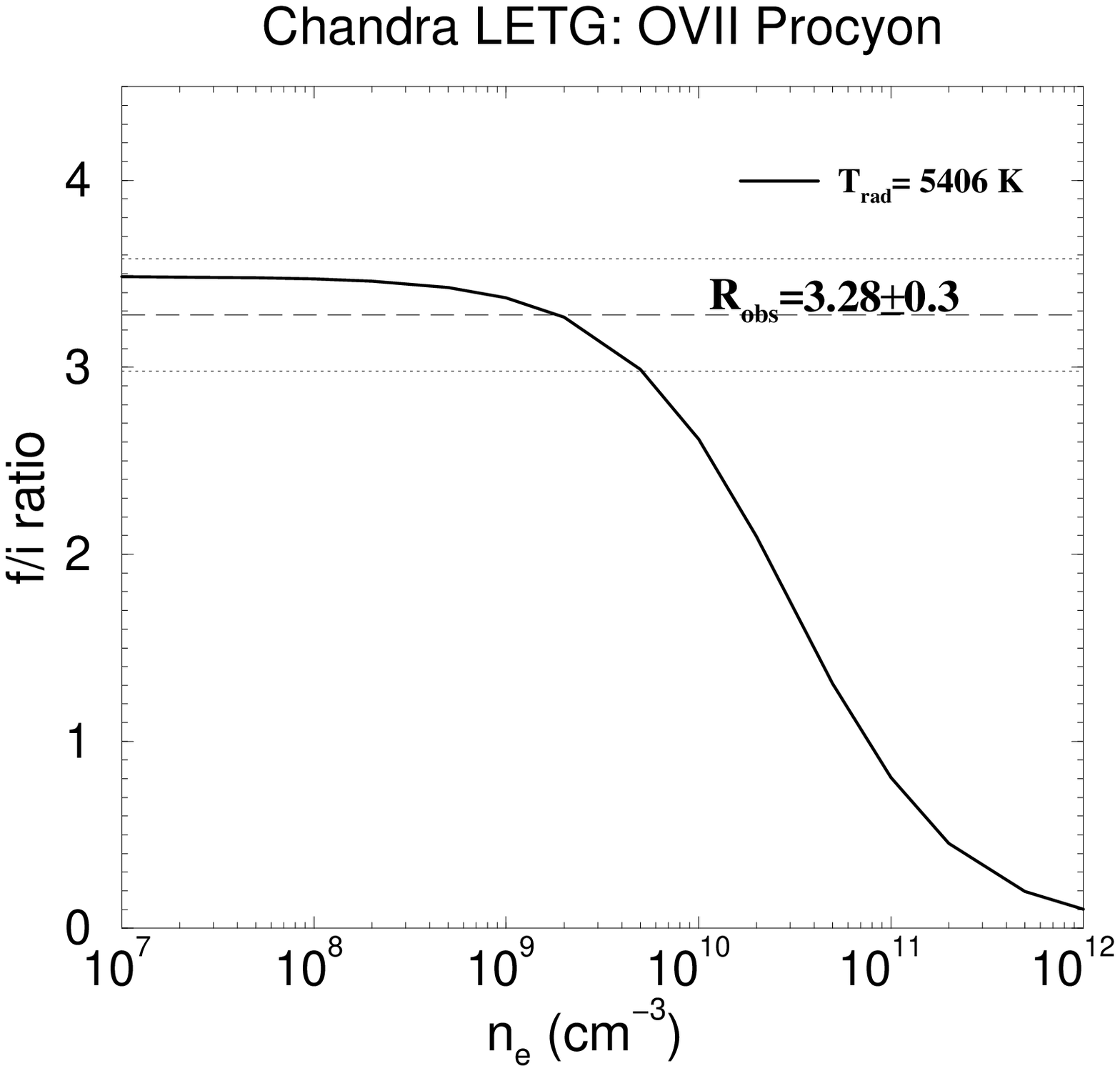}\includegraphics{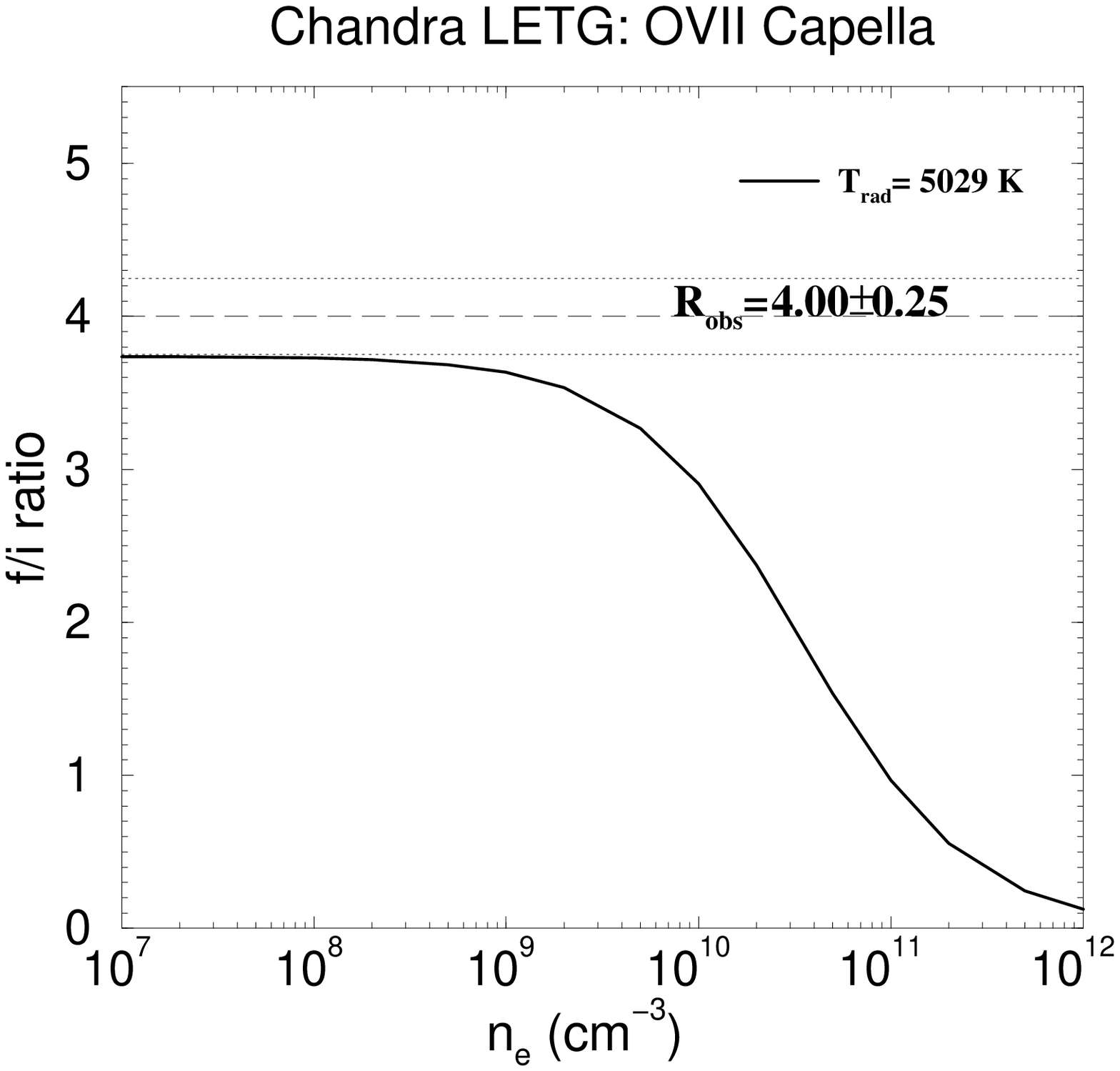}}
 \resizebox{\hsize}{!}{\includegraphics{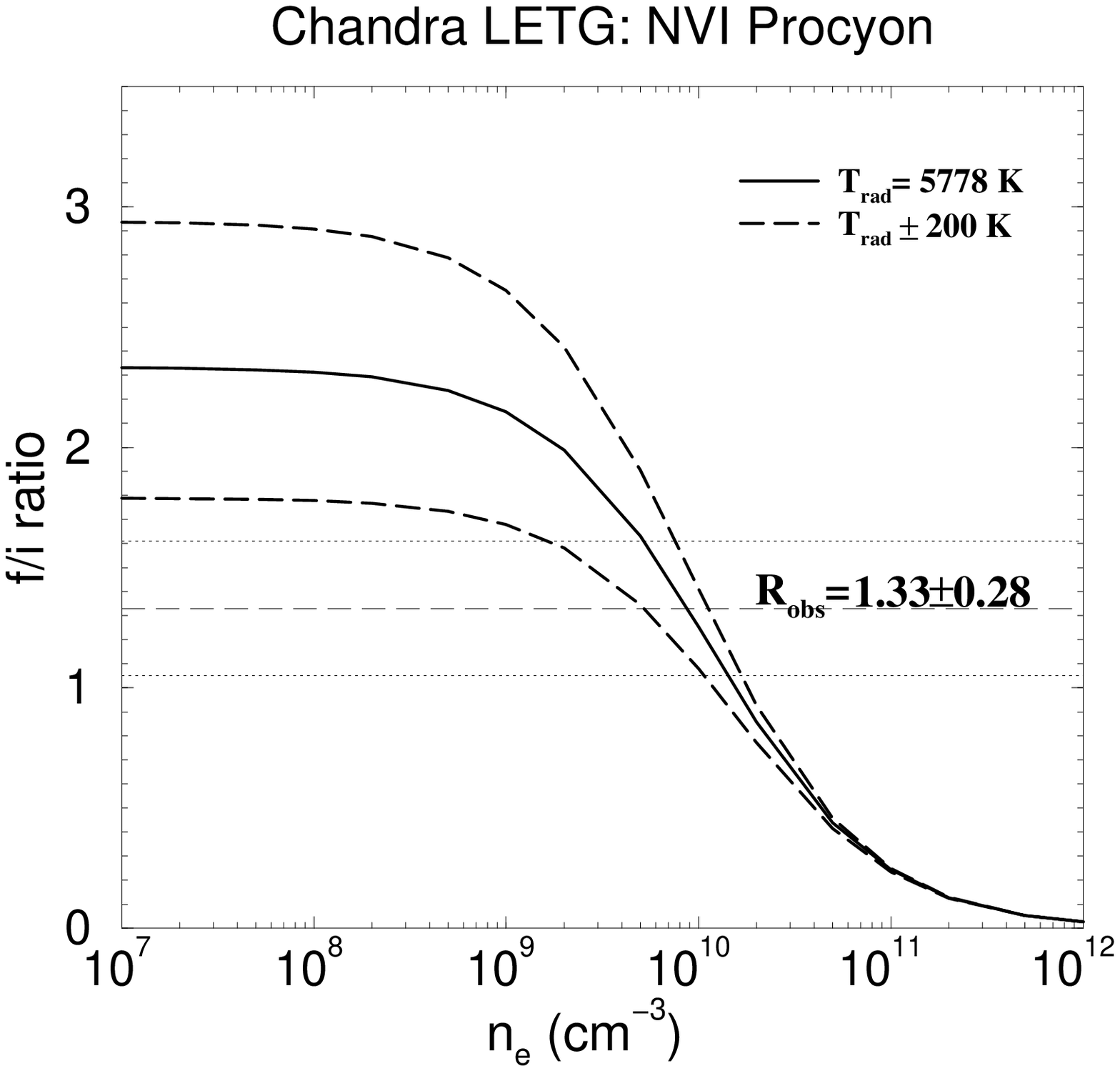}\includegraphics{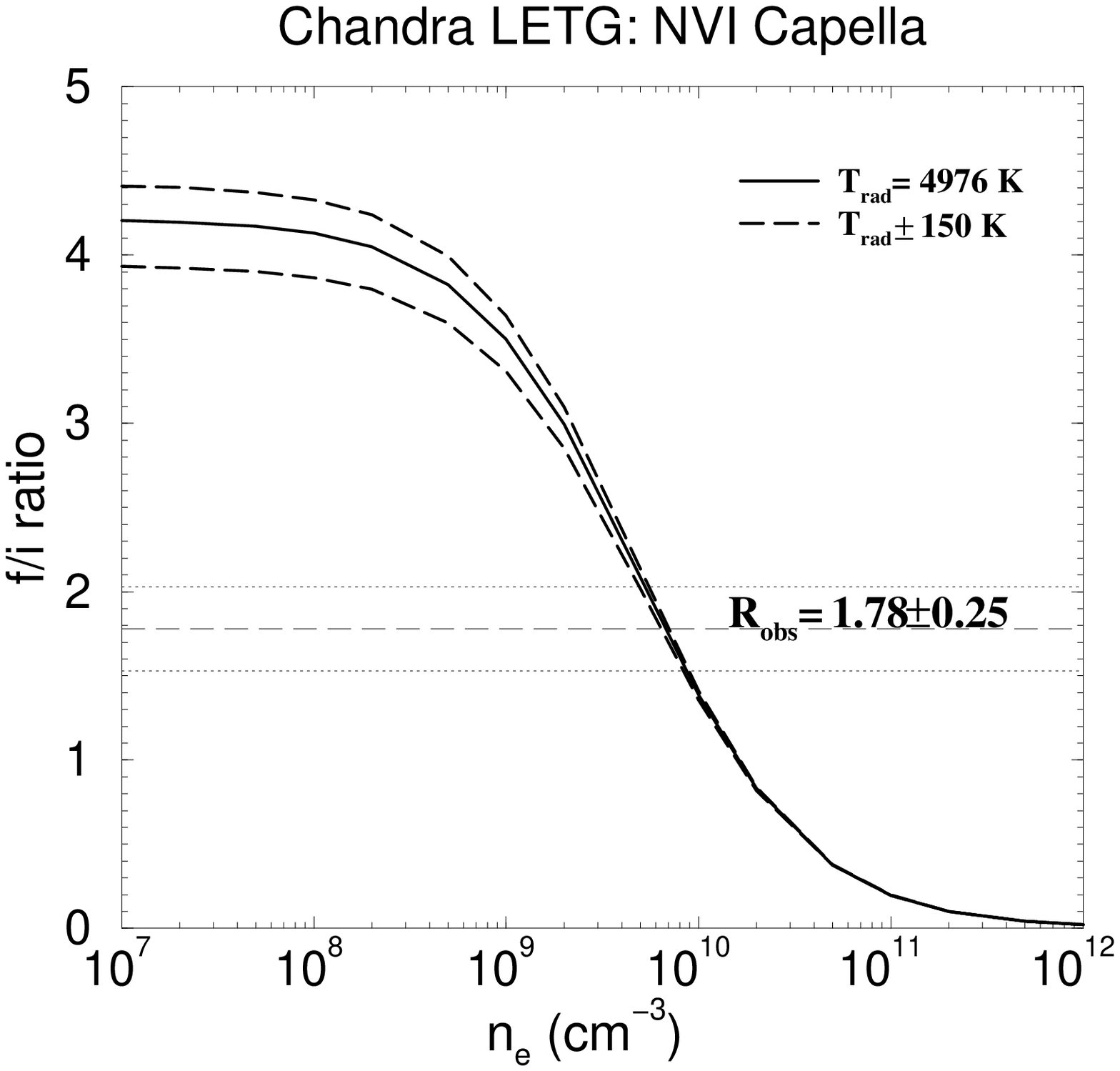}}
 \resizebox{\hsize}{!}{\includegraphics{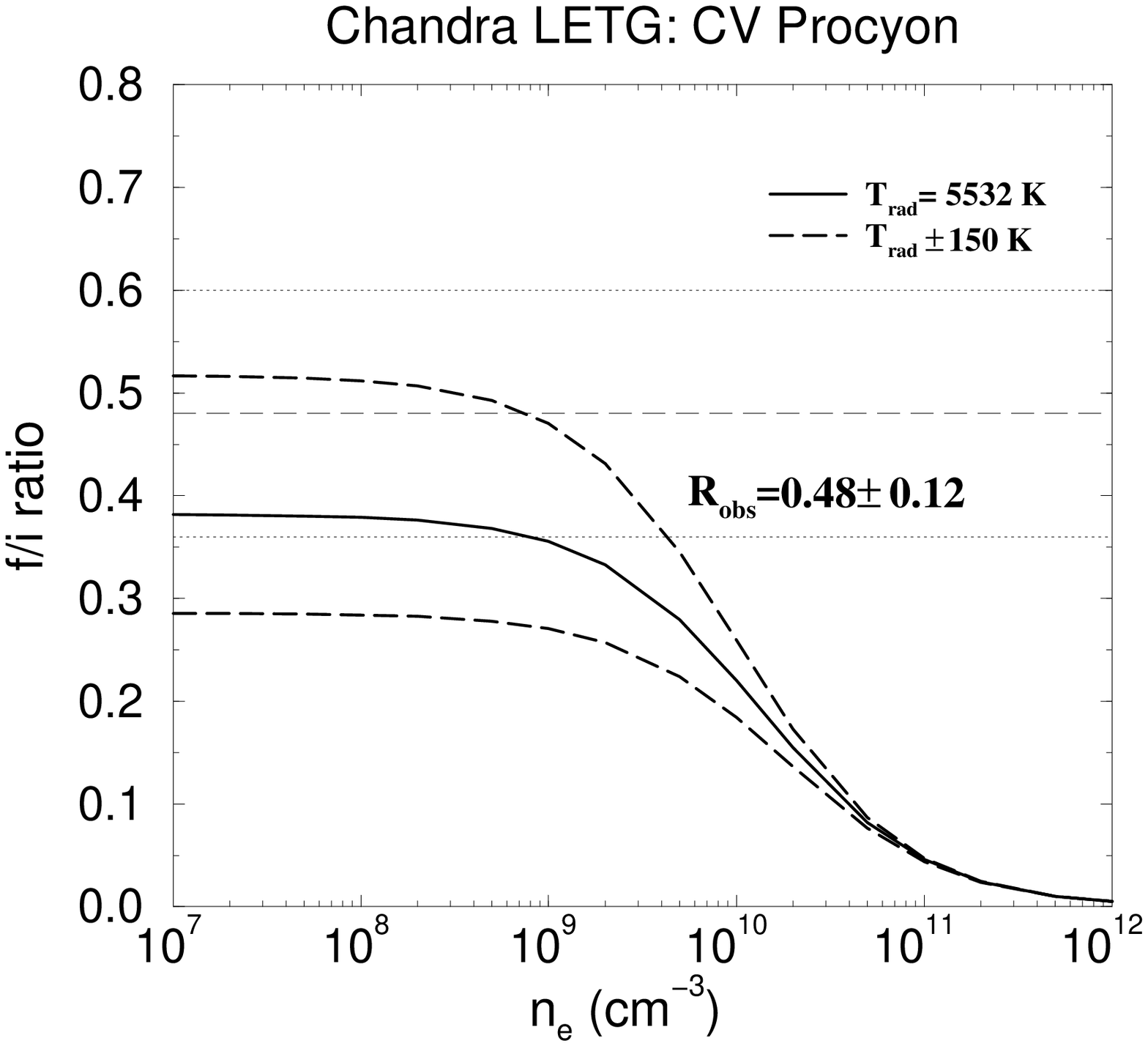}\includegraphics{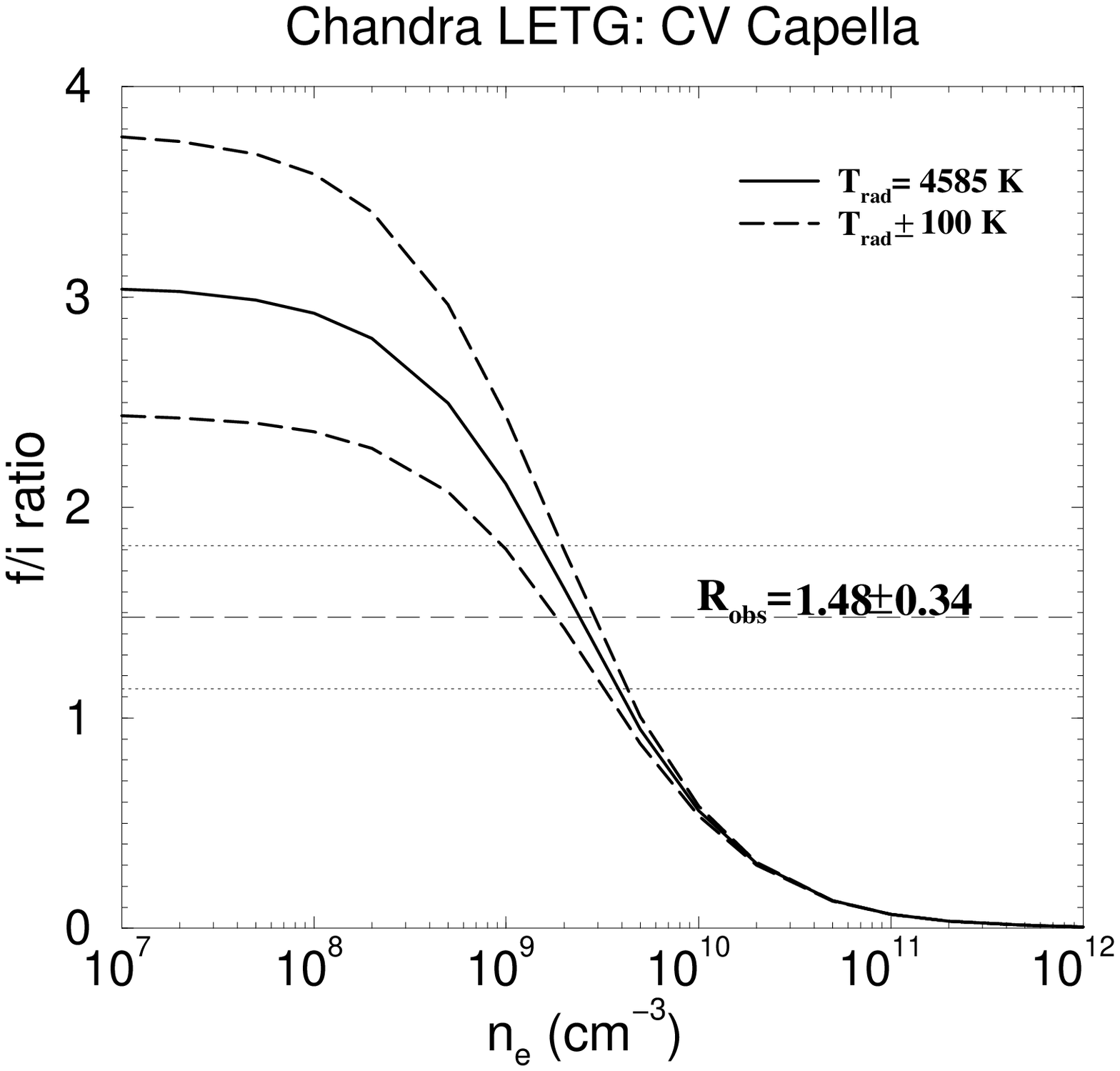}}
 \caption{ Theoretical curves in the radiation temperature range given in
Tab.~\ref{rad_temp} in comparison with the measured value of $R_{\rm obs}$
for the O\,{\sc vii}, N\,{\sc vi} and the C\,{\sc v} triplet for Capella and Procyon. The measured
R values from Tab.~\ref{tab_res} are corrected for detector efficiencies.}
 \label{dens}
\end{figure}

In Fig.~\ref{dens} we show for Procyon and Capella
the expected line ratio f/i as a function of the electron density $n_{\rm e}$
in comparison with the observed line ratio $R_{\rm obs}$ (corrected for detector
efficiencies) and its 1$\sigma$ error.
The expected curves are plotted for the radiation temperature range estimated
for the two stars as listed in Tab.~\ref{rad_temp}; for O\,{\sc vii} the
stellar radiation field does not significantly influence the f/i ratio as
expected. We used the formation
temperatures $T(G)$ calculated from the $G$ ratios as listed in
Tab.~\ref{ftemp}, thus assuming all the emission being produced at a single
temperature. In Tab.~\ref{dens_tab} we
summarize the derived densities and their 1$\sigma$ errors for Procyon and
Capella not accounting for errors in $T(G)$. 
No density values could be determined for the C\,{\sc v} triplet in Procyon
and for the O\,{\sc vii} triplet in Capella. Instead we give upper limits
for the two cases (Tab.~\ref{dens_tab}); for Capella only a 2$\sigma$
upper limit could be determined.\\
Obviously, for both stars the measured line ratios are within (Capella)
and very close to (Procyon)
the low-density limit $R_{\rm 0}$. At any rate, the measured f/i ratio is
larger for Capella than for Procyon, so one arrives at the somewhat unexpected
conclusion that the coronal density in the active star
Capella should be smaller than in the inactive star Procyon.\\

\begin{table}
\caption[ ]{\label{dens_tab}
Results of density diagnostics}
\begin{flushleft}
\renewcommand{\arraystretch}{1.2}
\begin{tabular}{r r r}
\hline
log $n_{\rm e}$/cm$^{-3}$&{\bf Capella}&{\bf Procyon}\\
\hline
C\,{\sc v}&$9.42\pm0.21$&$<$8.92\\
N\,{\sc vi}&$9.86\pm0.12$&$9.96\pm0.23$\\
O\,{\sc vii}&$<$9.38 &$9.28^{+0.4}_{-9.28}$\\
&$^{\mbox {\footnotesize (2$\sigma$ upper limit)}}$\\
\hline
\end{tabular}
\renewcommand{\arraystretch}{1}
\end{flushleft}
\end{table}

\subsection{Optical depth effects}
\label{optdepth}

In all of the above analysis we assumed all triplet lines to be optically thin.
In the following section we show this assumption to be consistent with our
results. Let us therefore assume that the optical depth in the resonance lines
is not small. This leads to a reduction of the measured resonance
line flux due to radiative scattering. At line center the optical depth is given
by the equation
\begin{equation}
\label{tau}
\tau=1.2\ 10^{-17}\left(\frac{n_i}{n_{el}}\right)A_z\left(\frac{n_H}{n_e}\right)\lambda\ f\sqrt{\frac{M}{T}}n_e\ell
\end{equation}
\noindent (\cite{schrijver94}) with the fractional ionization is denoted by
$n_i/n_{el}$, the elemental abundances by $A_z$,
the ratio of hydrogen to electron density is $n_H/n_e=0.85$,
the oscillator strength $f=0.7$ for all ions, the atomic number is $M$, the
wavelength is measured in \AA, the temperature in K, the electron density $n_e$
in cm$^{-3}$ and the mean free path is denoted by $\ell$.
In Eq.~\ref{tau} we adopt a value of unity for the fractional ionization and use
the solar abundances. We further assume $T$ at the peak line formation and
note, that $\tau$ is rather insensitive to the precise value of $T$. We can
determine -- for each resonance line -- that value of $n_e \ell$ which yields
an optical depth of unity. Adopting a maximum value of $n_e$ of $10^{10}$
cm$^{-3}$ (cf. Tab.~\ref{dens_tab}), we determine lower values of $\ell$ of
$1.3\,10^{12}$\,cm, $5.6\,10^{12}$\,cm and $6.0\,10^{11}$\,cm, for
O\,{\sc vii}, N\,{\sc vi} and C\,{\sc v}, respectively. Assuming a geometry most
suitable for resonance scattering, we can compute the respective emission
measures of
$n_e^2\ell^{\,3}$ respectively. Comparing these emission measures with those
derived from the measured line fluxes $f_\lambda$
\begin{equation}
EM=\frac{4\pi d^{\, 2} f_\lambda}{p_\lambda(T)}
\end{equation}
with the line cooling function $p_\lambda(T)$ and the distance $d$
(cf. Tab.\ref{star_prop}), shows the former to be larger by a few orders of
magnitude for both stars. This inconsistency shows that the assumption of a
non-negligible optical depth is invalid and we conclude that optical depth
effects are irrelevant for the analysis of He-like triplets in Procyon and
Capella.

\section{Conclusions}

\subsection{Comparison with the Sun}

Both Capella and Procyon are considered to be solar-like stars.
In the solar context the He-like triplets of oxygen, nitrogen and carbon have
been known for a long time and in fact the He-like ion density diagnostics have
been first developed to interpret these solar data. Many solar observations
are available for the O\,{\sc vii} triplet, while only very few observations have
been made for the N\,{\sc vi} and C\,{\sc v} triplets. The first observations of 
the C\,{\sc v} triplet are reported by \cite{aus66}, who obtained an f/i-ratio of 1.9,
while \cite{freeman70} found an f/i-ratio of 1.0. The latter authors also
observed the N\,{\sc vi} triplet, but failed to detect the intercombination line
and hence deduced f/i $>$ 1.9. \cite{brown86} observed the C\,{\sc v} and N\,{\sc vi} triplets
during a flare and found a mean value of f/i $\approx$ 2 with large scatter
between 0.1\,--\,4 around this mean for N\,{\sc vi} and a value of 0.21 (with a scatter
between 0.17\,--\,0.25) for C\,{\sc v}. In consequence, little can be said about the solar
N\,{\sc vi} line ratios because of the large measurement errors. The measured ratios for
both Capella and Procyon are certainly consistent
with the range of N\,{\sc vi} f/i values quoted by \cite{brown86}. As far as
C\,{\sc v} is concerned, the flare data yield very low f/i ratios indicative of
high densities. The C\,{\sc v} f/i ratio observed for Capella is higher, and that for
Procyon -- still higher than that observed for the \cite{brown86} flare --
we argue is due to large photospheric radiation fluxes. We therefore conclude
that the layers contributing the C\,{\sc v} emission in Capella and Procyon are at
lower density than those encountered in solar flares, and the same applies to
the layers emitting N\,{\sc vi}.

\begin{figure}
 \resizebox{\hsize}{!}{\includegraphics{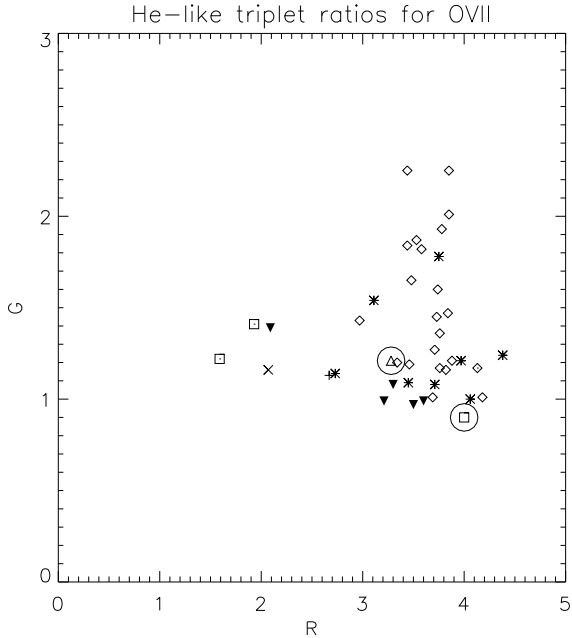}}
 \caption{\label{G_R}
Measurements of solar values for He-like ratios of O\,{\sc vii}:}
\mbox{\cite{freeman70} (dotted open box),}\\
\cite{mckenz82}: quiescent (diamond), flare (asterix),\\
\cite{mckenz78} (dotted diamond),\\
\cite{doyle80} (filled triangle upside down),\\
\cite{park75}: active region (plus symbol),\\
\cite{brown86}: flare (cross).\\
Encircled are our measurements of Procyon (open triangle) and Capella (open box).
\end{figure}

Observations of the O\,{\sc vii} triplet in the solar corona are reported by
\cite{freeman70}, \cite{mckenz78}, \cite{mckenz82}, all of which
refer to the quiescent corona, while some of the observations 
from \cite{mckenz82} and \cite{brown86} refer to flares and those
of \cite{park75} refer to active regions. In addition we considered
some data points collected from various sources cited by
\cite{doyle80}. In Fig.~\ref{G_R} we plot G vs. R of these solar observations
and compare these solar data with our {\it Chandra} measurements for
Procyon and Capella. For clarity we omitted individual error bars, which
are typically 0.1\,--\,0.3 for the solar measurements. As can be seen from 
Fig.~\ref{G_R}, most of the solar measurements yield R value between 3 and 4.
A few values are significantly lower around R $\approx$ 2, with all 
them referring to flares or active regions. Our {\it Chandra} measurements
(3.28 for Procyon, 4.0 for Capella) thus fall into the bulk part of the solar 
data. Unfortunately, the solar data are by no means unambiguous. It is 
not always
clear which regions the data refer to (except for the flare observations
by \cite{brown86}), and further, most of the observations are relatively old
and have not been taken with high spatial resolution; they are comparable
in some sense to our full disk observations of stellar X-ray sources.
In addition, the instrumentation used in
different experiments is quite different and each experiment is affected
by its own specific difficulties. Nevertheless it seems fair to state
that most O\,{\sc vii} measurements referring to quiescent conditions are close
to the low density limit, thus indicating that the physical properties 
of the O\,{\sc vii} emitting layers
in Capella and Procyon
should not be that different from those in the Sun; a little
puzzling in this context is the large scatter of the solar G data.
In contrast, the solar flare data (R $\sim$ 2) is much lower than any of the
other measurements, consistent with a high density plasma.

\subsection{Comparison between Procyon and Capella}

Inspection of Tab.~\ref{dens_tab} suggests that $n_{Procyon} \ge n_{Capella}$
for the densities derived from C\,{\sc v}, N\,{\sc vi}, and O\,{\sc vii}; there is definitely no
evidence for $n_{Procyon} < n_{Capella}$. These findings are in
contradiction to the claims made by \cite{dupree93}, who argue on the
basis of Fe\,{\sc xxi} line ratios measured with the EUVE satellite that 
$n_{Capella} \sim 10^{13}$ cm$^{-3}$. A recent analysis of the
long wavelength part ($\lambda > 90$~\AA) of the LETGS spectrum of
Capella by \cite{mewe00b}, however, yielded plasma densities of
$5 \times 10^{12}$ cm$^{-3}$ at most for the plasma emitting in the
Fe\,{\sc xxi} lines with no evidence for any deviations from the low density
limit. Of course, these values refer to
those plasma layers emitting in the Fe\,{\sc xxi} line, which ought to be at
far higher temperature ($\approx$ 8 MK) than
those emitting in the C\,{\sc v}, N\,{\sc vi}, and O\,{\sc vii} lines (1\,--\,2 MK).
If we assume all the observed emissions to come from the same 
magnetically confined structures (loops), one expects the gas pressure
to be more or less constant and hence the density at lower temperatures even
higher. However, plasma densities (for Capella) of
$\approx 10^{13}$ cm$^{-3}$ at temperatures of 2 MK are clearly inconsistent
with our LETGS spectra. Therefore the plasma emitting in the Fe\,{\sc xxi} line
is either at the low density limit or the Fe\,{\sc xxi} emission comes from
different structures than the emissions from He-like ions.

Our temperature measurements of the line forming regions (cf.,
Tab.~\ref{ftemp}) are not fully conclusive, although they suggest that the same
lines of C\,{\sc v}, N\,{\sc vi}, and O\,{\sc vii} respectively are
formed at a somewhat higher temperature (at most a factor of 2) 
in Capella as compared to Procyon. As a consequence we must have 
$p_{Procyon} \approx p_{Capella}$ for the gas pressures in the respective
line forming regions. 
Capella's peak coronal temperature is clearly much higher than 
Procyon's.
If we again assume -- going along with the solar analogy -- that for both
Procyon and Capella (all of) the X-ray emission originates from magnetically 
confined plasma loops, the loop top temperatures in Capella must be
higher than in Procyon. The loop scaling laws then imply for the
typical loop length scales $L$ for Procyon and Capella
$\frac {L_{Capella}} {L_{Procyon}} = 
(\frac {T_{max,Capella}}{T_{max,Procyon}})^3 $. A conservative estimate
yields $\frac {T_{max,Capella}}{T_{max,Procyon}} > 2$, thus
${L_{Capella}} \ge 8~{L_{Procyon}}$. The conclusion then appears
inescapable that the typical sizes of the magnetic structures in
Procyon and Capella are quite different by probably at least one order
of magnitude.

Next, it is instructive to compare the total mean surface fluxes emitted in
the helium- and
hydrogen-like ions of carbon, nitrogen and oxygen for Procyon and Capella.
Using the measured count ratios and correcting for the effects of distance,
exposure times and the surface areas of the
stars we find for the C\,{\sc v}, C\,{\sc vi}, N\,{\sc vi}, N\,{\sc vii},
O\,{\sc vii}, and O\,{\sc viii} lines values of (Capella)/(Procyon) of 0.96,
1.37, 1.09, 4.9, 1.87, and 9.71 respectively. Therefore up to those 
temperatures where the C\,{\sc v}, C\,{\sc vi}, and N\,{\sc vi} lines are
formed, the mean surface fluxes in the two stars hardly differ at all;
pressures and temperatures are also quite similar. We therefore conclude
that the physical characteristics and global properties (such as filling factor)
of Procyon's and Capella's corona at a level of $T \sim 10^6$ K are very similar.

It is interesting to again perform a comparison to the Sun.
\cite{freeman70} quote -- for their SL801 rocket flight on Nov 20 1969 --
resonance line fluxes of 4.2 10$^{-3}$, 2.1 10$^{-3}$, and 3.1 10$^{-3}$
erg/(cm$^2$\,sec) for O\,{\sc vii}, N\,{\sc vi}, and C\,{\sc v} respectively.
Correcting for the distance between Earth and Sun we find average surface
fluxes of 200, 100, and 150 erg/cm$^2$/sec for O\,{\sc vii}, N\,{\sc vi},
and C\,{\sc v} respectively on the
solar surface. Carrying out the same calculation for our {\it Chandra}
data, we find -- using the effective areas given in Tab.~\ref{eff_ar} --
mean surface fluxes in O\,{\sc vii} of 3700 and 1980 erg/(cm$^2$\,sec),
in N\,{\sc vi} of 480 and 430 erg/(cm$^2$\,sec),
and in C\,{\sc v} of 830 and 860 erg/(cm$^2$\,sec)
for Capella and Procyon respectively. We therefore recover our
previous finding that Capella's and Procyon's surface fluxes are quite 
similar, and determine in addition that both stars exceed typical (?)
solar values by one order of magnitude. Since the physical conditions of the
line emitting regions are quite similar, we conclude that the coronal
filling factors are larger.

Evaluating now the coronal pressure for Procyon we find -- using the 
densities and temperatures derived from
N\,{\sc vi} -- $p$(N\,{\sc vi}) = 4.4~dyn/cm$^2$, a value also very
typical for solar active regions. From the loop scaling law (\cite{RTV78})
$T_{max} = 1.4 \times 10^3 (p L)^{(1/3)}$ we deduce a typical length 
of $2.7 \times 10^8$ cm using $T_{max} = 1.5 \times 10^6$~K.
This value must be considered as a lower limit to the
probable length scale because of the unfortunate sensitive dependence of $L$ on
$T_{max}$ through $L \sim T_{max}^3$; $T_{max}$ is likely somewhat higher than
$T_{max} = 1.5 \times 10^6$~K as suggested by the temperatures derived from the
O\,{\sc viii} data. At any rate, however, the conclusion is, that the X-ray
emission originates from low-lying loops and hence also our {\it Chandra} data
support the view that Procyon's corona has an appearance very similar to the 
Sun's, just like
the conclusion drawn by \cite{schmitt96b} from their EUVE spectra. It then
follows that the sizes of the magnetic loops in Procyon's corona are similar
to the loops found in the solar corona. 
Adopting -- for argument's sake --
a characteristic length of L$_{Proc}$ = 1$\times 10^9$ cm, we then find
for Capella $L_{Capella} > 8 \times 10^{9}$ cm, and possibly even 
a significant 
fraction of Capella's radius. This result is somewhat puzzling. On the Sun,
loop structures, i.e., magnetically closed topologies, are occasionally
found at larger heights, but they are always of low density and 
they never contribute significantly to the 
overall X-ray emission. It will be interesting to see whether this
behavior is typical for active stars in general, or whether it applies
only to Capella. After all, Capella may not be the prototypical active star.
Its corona is not as hot as that of other stars, it does not produce flares
and its radio emission is very weak.

\begin{acknowledgements}
J.-U.\,N. acknowledges financial support from Deutsches Zentrum f\"ur Luft- und
Raumfahrt e.V. (DLR) under 50OR98010.\\
The Space Research Organization Netherlands (SRON) is supported financially by NWO.
\end{acknowledgements}

%#######################################################
%               BIBILIOGRAPHY
%#######################################################

\end{document}